\documentclass[pra,showpacs,letterpaper,showpacs,superscriptaddress,nofootinbib]{revtex4-1}
\usepackage{graphicx,amsmath,amssymb,amsfonts,latexsym,color,dcolumn,bm,subfigure}
\usepackage{mathrsfs,natbib}

\newcommand{\p}{{\rm p}}
\newcommand{\s}{{\rm S}}

\newcommand{\gp}{g_{\rm pr}}
\newcommand{\ga}{g_{\rm D}}

\begin{document}

\title{Engineering dissipation with phononic spectral hole burning}

\author{R. O. Behunin}
\affiliation{Department of Applied Physics, Yale University, New Haven, Connecticut 06511, USA}
\author{P. Kharel}
\affiliation{Department of Applied Physics, Yale University, New Haven, Connecticut 06511, USA}
\author{W. H. Renninger}
\affiliation{Department of Applied Physics, Yale University, New Haven, Connecticut 06511, USA}
\author{P. T. Rakich}
\affiliation{Department of Applied Physics, Yale University, New Haven, Connecticut 06511, USA}

\date{ \today}

\begin{abstract}
Optomechanics, nano-electromechanics, and integrated photonics have brought about a renaissance in phononic device physics and technology \cite{Kippenberg2008,Regal2008,Bochmann2013,Mahboob2013}. Central to this advance are devices and materials that support ultra long-lived photonic and phononic excitations, providing access to novel regimes of classical and quantum dynamics based on tailorable photon-phonon coupling \cite{Schliesser2009,Lee2012}. Silica-based devices have been at the forefront of such innovations for their ability to support optical excitations persisting for nearly 1 billion cycles \cite{Lee2012}, and for their low optical nonlinearity. Remarkably, acoustic phonon modes can persist for a comparable number of cycles in crystalline solids at cryogenic temperatures \cite{Goryachev2013d,Meenehan2015a}, permitting radical enhancement of photon-phonon coupling. However, it has not been possible to achieve similar phononic coherence times in silica, as silica becomes acoustically opaque at low temperatures \cite{Phillips1987}. In this paper, we demonstrate that intrinsic forms of phonon dissipation are greatly reduced (by $>$ 90\%) using nonlinear saturation with continuous driving fields of disparate frequencies.
Furthermore, we demonstrate steady-state phononic spectral hole burning for the first time, and show that this technique for controlling dissipation in glass produces a wide-band transparency window. These studies were carried out in a micro-scale fiber waveguide where the acoustic intensities necessary to manipulate phonon dissipation can be achieved with optically generated phonon fields of modest (nW) powers. Combining experiment and theory, we developed a simple model that explains both dissipative and dispersive changes produced by phononic saturation. In showing how the dissipative and dispersive properties of glasses can be manipulated using external fields, we open the door to dynamical phononic switching and the use of glasses as low loss phononic media which may enable new forms of controllable laser dynamics, information processing, and precision metrology. 
\end{abstract}


\maketitle

\section{Introduction}

In recent years, a range of micro- and nano-scale systems have emerged that permit engineerable coupling between photons and acoustic phonons. The strength of this nonlinear light-matter coupling is greatly enhanced at low temperatures where intrinsic phonon dissipation becomes exceedingly small, producing radically enhanced phonon coherence (or lifetimes). For instance, 
at cryogenic temperatures phonon coherence times are increased by factors of $10^3-10^6$ over their room temperature values at cryogenic temperatures \cite{Goryachev2013d,Meenehan2015a}, making new regimes of optomechanical and electromechanical phenomena accessible. Applications include quantum and classical information processing, controllable laser dynamics, and high-precision sensing 
\cite{Shin2013,Shin2015,Chan2011,Fiore2011,Weis2010,Goryachev2013d,Cheung2016,Regal2008,Lee2012,Kang2009,DelHaye2007,Kippenberg2011}. However, in amorphous media (glasses) this picture becomes more complex. While silica glass is an exceptional optical medium \cite{Lee2012}, its complex elastic properties give rise to nonlinear phonon dynamics that can destroy phonon coherence even at cryogenic temperatures \cite{Phillips1987}.

At low temperatures, the dominant form of phonon dissipation in amorphous media originates from an ensemble of acoustic `atoms' that absorb and emit phonons. These acoustic atoms are conventionally modeled as strain-active two-level systems (TLS) that can be driven into transparency (or saturated) with high-intensity phonon fields \cite{Phillips1987}. Hence, similar to microwave and optical spectral hole burning in solid state systems \cite{Portis1953,Szabo1975,Jessop1980,Basche1992}, phononic spectral hole burning can also be achieved in glasses at low temperatures. Understanding the underlying dynamics of this process makes it possible to radically enhance phonon lifetimes by driving acoustic atoms into transparency. The seminal studies by Golding {\it et al.} \cite{Golding1973,Golding1976b}, show that silica can be made highly acoustically transparent at low temperatures with high driving fields. These experiments suggest \cite{Golding1976b}  that ultra-long lived phonon modes (Q-factors of $10^6$ at 0.6 GHz) can be achieved despite the large intrinsic dissipation of glass. By harnessing this nonlinear dynamics, it may be possible to dramatically enhance the performance of glass-based optomechanical and electro-mechanical systems to enable new regimes of laser dynamics, quantum-limited detection, and precision metrology.

At present, the ultimate limits of phonon dissipation in glass are not fully understood. Few experiments have investigated saturable acoustic dissipation in glass, and the broadband effects of continuous driving on the dissipative and dispersive properties of elastic media is unexplored at cryogenic temperatures \cite{Golding1973,Golding1976b,Arnold1974,Arnold1975,Arnold1978}.
Moreover, an array of complex cross relaxation processes, such as spectral diffusion and phonon assisted energy transfer, can dramatically influence the material dynamics of glass \cite{Arnold1975,Arnold1978,Black1977}. 
The obscure nature of the microscopic couplings that underlie these processes impede theory-only efforts to accurately describe spectral hole burning. Hence,  closely coupled theoretical and experimental efforts are necessary to develop a useful framework to understand phononic spectral hole burning.

In this paper, we examine phononic spectral hole burning in silica at 9 GHz frequencies using non-invasive optical probes at cryogenic temperatures. 
These measurements reveal that under steady-state driving, acoustically opaque glasses can be driven into transparency over a wide spectral band.
Combining these saturation measurements with experimental sound velocity information we construct a more complete picture of the system's non-equilibrium response. To understand these data we formulate a new model to describe the non-equilibrium dynamics of driven glasses (ensembles of acoustic atoms). 
We show that this model, which is built on established models of glass physics, explains the dispersive and dissipative changes produced by phononic spectral hole burning. 
Using parameters obtained from prior studies \cite{Behunin2016b} we find good agreement between this model and our spectral hole burning data. 
We show that a phenomenological treatment accounting for resonant energy transfer and spectral diffusion adequately describes the dynamics of spectral hole burning.
Furthermore this model suggests that modest driving fields (nW) can produce transparency over a multi-GHz bandwidth, comparable in magnitude to the drive frequency, turning otherwise acoustically opaque glasses into highly transparent media.

\begin{figure*}[t!]
\begin{center}
\includegraphics[width=\textwidth]{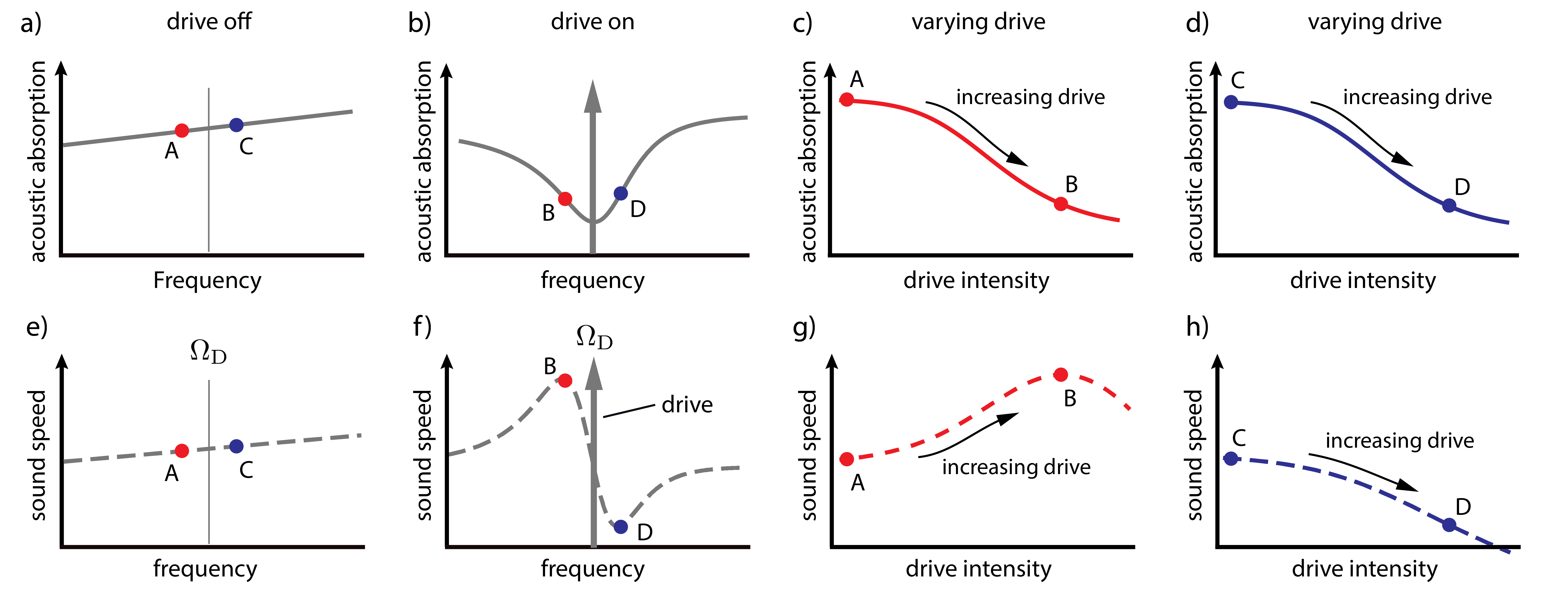}
\caption{Modification of elastic properties by phononic spectral hole burning. 
Acoustic absorption of a) an undriven and b) a driven system. b) Absorption is saturated in a band of frequencies centered on an intense drive field. Saturation of acoustic absorption for frequencies c) below and d) above a variable intensity drive. Sound speed in e) an undriven and f) a driven system. f) Large dispersive shifts accompany changes in the absorptive properties caused by driving, as mandated by the Kramers-Kronig relations. An intense drive shifts the sound speed in opposite directions for frequencies g) below and h) above the drive.   
}
\label{cartoon}
\end{center}
\end{figure*}

The salient information obtained through spectral hole burning measurements is summarized in Fig. \ref{cartoon}. Here, we consider the effect of a strong phonon drive field at frequency $\Omega_{\rm D}$, which drastically alters the populations of the acoustic absorbers, on the elastic properties of a medium. Once the acoustic absorbers are driven into a nonequilibrium
steady-state, a weak variable-frequency probe field measures the absorptive and dispersive properties of the system without altering the population.
For low drive intensities the acoustic absorbers remain in thermal equilibrium, and the probe measures the intrinsic elastic properties (Figs. \ref{cartoon}a \& e). Similar to crystalline solid state  or atomic systems \cite{Portis1953,Szabo1975,Jessop1980,Siegman1986,Basche1992}, an intense drive saturates acoustic dissipation in a band of frequencies (Fig. \ref{cartoon}b). The suppression of absorption with increased drive intensity at distinct frequencies from the drive signifies the formation of a spectral hole (Fig. \ref{cartoon}c \& d). 

Dispersive measurements at these same probe frequencies permit us to construct a more complete picture of the population changes at nearby frequencies. Kramers-Kronig relations interlink changes in absorption to shifts in sound speed. Therefore, a strong drive reshapes acoustic dispersion as well (see Fig. \ref{cartoon}f). From Fig. \ref{cartoon}f, it is apparent that changes in sound speed produced by the drive-field depend sensitively on the frequency of measurement; saturation of the absorption produces either an increase or decrease in the sound velocity for frequencies above and below the drive (Fig. \ref{cartoon}g \& h).  
We will show that measurements of these nonlinear phonon dynamics at various frequencies yield information about the spectrum, coupling, density, and dynamics of the acoustic atoms in a system. This information lays the foundation for an empirically motivated theory of phononic spectral hole burning which in turn establishes the limits of phonon coherence in amorphous media. 

Established theories of acoustic dissipation in glasses suggest that the physics underlying spectral hole burning in amorphous media involves a number of complex processes (see Supplemental Information Sec. A). 
TLS-based microscopic models lead us to expect the phenomena of phononic spectral hole burning in low temperature glasses to share many basic characteristics with widely studied forms of microwave, optical, and persistent spectral hole burning \cite{Portis1953,Szabo1975,Jessop1980,Basche1992,Skinner1996,Jankowiak1987,Bjorklund2012}. At root, phononic hole burning comes about by populating the excited states of acoustic TLS with a strong drive, which in turn renders these TLS transparent. Subsequently, a diversity of cross-relaxation \cite{Siegman1986} processes transfer this energy throughout the ensemble of acoustic atoms. This is important because these energy transfer processes determine the bandwidth of the spectral hole. 

While optical spectral hole burning in crystals has many common features, there are some aspects of energy transfer in low temperature glasses that are unique. For instance, acoustic TLSs lack an intrinsic energy scale, having individual energy levels distributed over a broad, almost white, spectrum. Hence, in contrast with an inhomogeneously broadened ensemble of resonant absorbers, there's no limit, in principle, to how far (spectrally) energy transfers from the drive by absorption and reemission. Additionally, acoustic atoms interact directly with each other through the static elastic field. These properties enable cross relaxation processes such as quasi-resonant emission and absorption, spectral diffusion, and phonon-assisted energy transfer to cascade energy from the drive to neighboring frequencies (see Supplemental Information Sec. A). 

It is known that direct interactions among acoustic atoms play a key role in dephasing through spectral diffusion \cite{Arnold1975,Arnold1978}. However, the nature and strength of these couplings are still not fully understood. Indeed, recent observations imply that these direct interactions can be quite strong \cite{Burnett2014a}, raising questions about the relative importance of higher order processes in phononic spectral hole burning. In what follows, we address these challenges by combining new forms of nonlinear phonon spectroscopy with phenomenologically based TLS models to estimate the bandwidth scaling of spectral hole burning, and to develop a useful description of steady-state phononic spectral hole for application to broad classes of disordered materials.

\begin{figure*}
\begin{center}
\includegraphics[width=\textwidth]{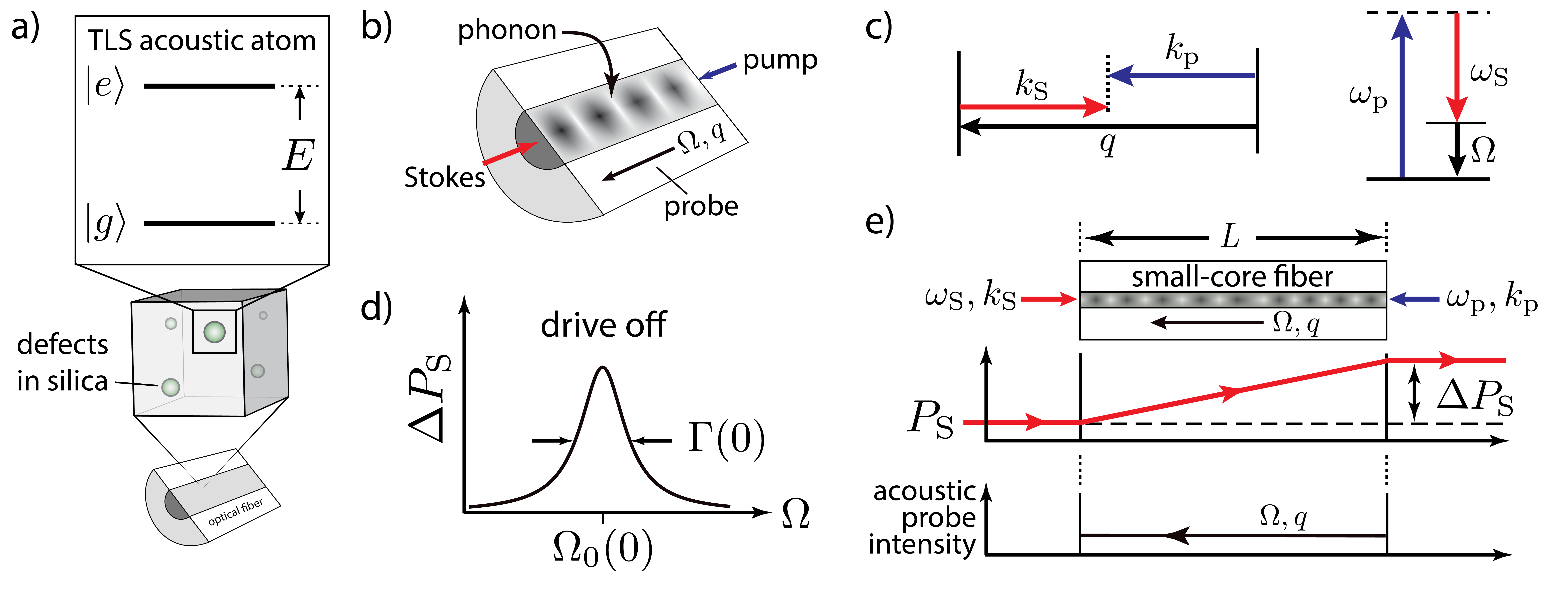}
\caption{a) Small core fiber system to measure phononic spectral hole burning: acoustic atoms reside in fiber core. b) Brillouin scattering in a fiber system: counter propagating optical pump and Stokes fields generate guided phonons in the fiber core, c) phase matching conditions for Brillouin scattering. d) Energy transfer $\Delta P_\s$ as a function of pump-Stokes detuning $\Omega$, giving phonon frequency and dissipation rate. e) Pictorial representation of energy transfer occurring in system. 
}
\label{systemSketch}
\end{center}
\end{figure*}

{\it  System, Experiment \& Brillouin Scattering:}
We perform nonlinear phonon spectroscopy on silica glass at cryogenic temperatures using a noninvasive form of nonlinear optical spectroscopy. Laser light is used to generate and detect phonons through photoelastically mediated optomechanical (or Brillouin) coupling. 
In order to enhance this light-sound coupling we use small core silica-based optical fiber 
to perform spectral hole burning studies (Fig. \ref{systemSketch}a). The germanium doping in the core of the optical fiber provides simultaneous guidance of tightly confined light and sound fields;
tight confinement of both light and sound within the fiber greatly enhances the photoelastic (Brillouin) coupling. Using this fiber waveguide as our specimen, injected light beams can detect phonons in the core of this fiber, and modest optical powers can generate high-intensity acoustic fields (see Fig. \ref{systemSketch}b). 

We explore the nonlinear acoustic properties of silica glass (within the fiber core) through Brillouin-based phononic pump-probe measurements. Through backward stimulated Brillouin scattering two counter-propagating optical waves couple to a (narrow band) traveling phonon mode. 
This coupling mediates energy transfer from a high frequency (pump) optical wave to a lower frequency (Stokes) optical wave, of respective frequencies ($\omega_\p$, $\omega_\s$) and wavevectors ($k_\p$, $k_\s$), while simultaneously generating phonons (of frequency $\Omega$ and wavevector $q$) in the fiber core.

Brillouin-mediated energy transfer between the pump and the Stokes waves provides information about the phonon generation rates as well as the dissipative and dispersive phononic properties of the medium.
Since each time a phonon is generated, a Stokes photon is also emitted, the power transfer between the pump and Stokes beams provides a direct measure of the phonon generation rates within the specimen. This Stokes energy transfer is maximized when phase matching between the wave frequencies ($\omega_\p = \omega_\s + \Omega$) and wavevectors ($k_\p = -k_\s + q$) are satisfied (Fig. \ref{systemSketch}c). In our fiber waveguide this resonance condition is satisfied for the phonon frequency given by $\Omega_0 \approx (2 n v/c) \omega_\p$ where $c/n$ and $v$ are the phase velocities of light and sound, and $n$ is the index of refraction. Hence, changes in sound velocity are encoded in the resonance frequency ($\Omega_0$) of the energy transfer.
 Through Brillouin coupling, the Stokes field power ($P_\s$) is increased by an amount $\Delta P_\s$, yielding an energy transfer spectrum of the form \cite{Boyd2003}
\begin{equation}
\label{BGS}
\Delta P_\s \approx \frac{(\Gamma(J_{\rm D})/2)^2}{(\Omega_0(J_{\rm D}) -\Omega)^2 + (\Gamma(J_{\rm D})/2)^2} G_B P_\p P_\s L.
\end{equation}
Here, $P_\p$ is the power of the pump, $L$ is the length of the fiber segment, $\Gamma(J_{\rm D})$ is the acoustic decay rate, and $\Omega = \omega_\p-\omega_\s$ is the frequency of the emitted phonons as required by energy conservation. Note that $G_B$ is proportional to $1/\Gamma(J_{\rm D})$ and quantifies the strength of Brillouin scattering (see Supplemental Information Sec. B). The energy transfer ($\Delta P_\s$ of Eq. \ref{BGS}) is a Lorentzian with a full-width at half maximum that coincides with the phonon dissipation rate $\Gamma(J_{\rm D})$. Hence, measurements of energy transfer as a function of the optical detuning $\Omega = \omega_\p-\omega_\s$, as depicted in Fig. \ref{systemSketch}d, provide a measure of both the sound speed (encoded in the resonant frequency $\Omega_0$) and acoustic absorption (encoded in the energy transfer bandwidth $\Gamma$).    

\begin{figure*}
\begin{center}
\includegraphics[width=\textwidth]{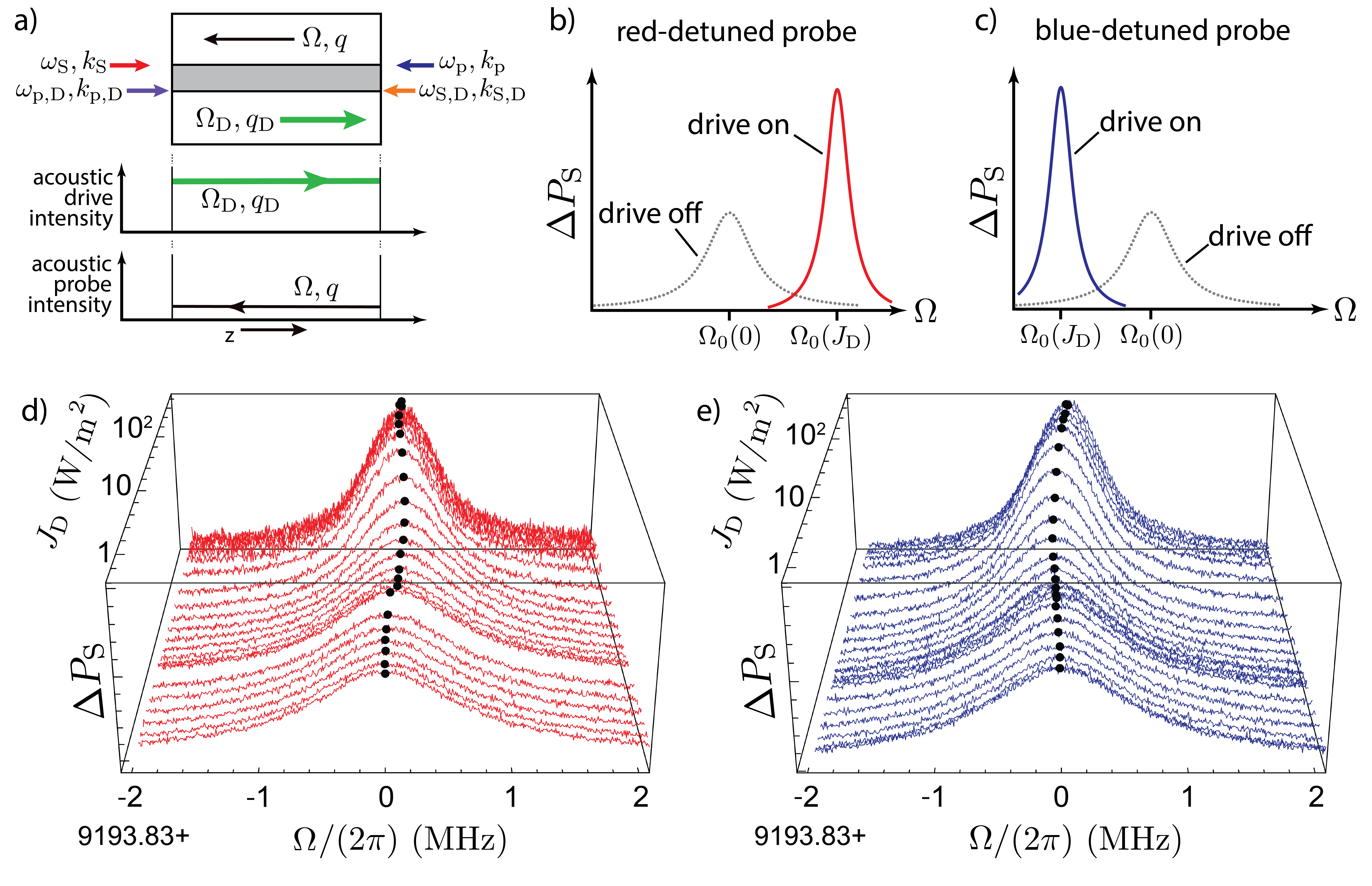}
\caption{ 
a) To investigate phononic spectral hole burning an intense drive and a counter propagating probe are generated in the fiber core with four distinct optical fields. 
The intense drive distinctly modifies the spectrum of scattered probe light $\Delta P_{\rm S}$ for a b) red-detuned acoustic probe and c) a blue-detuned acoustic probe. We observe these behaviors in measurements of $\Delta P_\s$ for a d) red-detuned and e) blue-detuned probe as the drive intensity $J_{\rm D}$ is varied. Black dots mark peak frequency $\Omega_0(J_{\rm D})$ for each Brillouin spectrum.}
\label{systemSketch-2}
\end{center}
\end{figure*}

To quantify the nonlinear phonon dynamics of our glass system, we use two optical drive frequencies ($\omega_{\p,{\rm D}}$ and $\omega_{\rm S,D}$) to generate an acoustic drive ($\Omega_{\rm D}, q_{\rm D}$) of variable intensity $J_{\rm D}$ propagating in the $+z$-direction (Fig. \ref{systemSketch-2}a). This drive leads to phononic spectral hole burning that modifies the sound speed (encoded in $\Omega_0(J_{\rm D})$) and absorption ($\Gamma(J_{\rm D})$) (Fig. \ref{systemSketch-2}b \& c). With fixed drive frequency and intensity, we measure the dissipative and dispersive properties of the glass using low intensity probe beams at distinct optical frequencies ($\omega_{\p}$ and $\omega_{\rm S}$) that couple to phonons ($\Omega, q$) propagating in the $-z$-direction (see Supplemental Information Sec. C for experimental details). 

Changes to the phonon dissipation and dispersion due to an intense drive are observed from stimulated Brillouin spectra of the type seen in Fig. \ref{systemSketch-2}d \& e. 
These spectra are obtained by measuring the energy transfer from pump to Stokes as the relative detuning $\Omega$ is swept through the resonant frequency. They contain information about the phonon generation rate (peak height), as well as the sound speed ($\Omega_0(J_{\rm D}$)), and the acoustic absorption (full width at half max $\Gamma(J_{\rm D})$). 
These data reveal several features of phononic spectral hole burning. 
Brillouin energy transfer and the phonon lifetime are both enhanced as the intensity of the drive is increased, consistent with the saturation of acoustic dissipation by the drive. In addition, the peak frequency shifts with intensity in a direction determined by the pump-drive detuning, indicating the presence of a spectral hole (Figs. \ref{cartoon} \& \ref{systemSketch-2}). 

{\it Background Theory:} 
To understand these data, we develop a simple phenomenological model of phononic spectral hole burning that demonstrates good agreement with our nonlinear frequency shift and dissipation measurements (Fig. \ref{systemSketch-2}d \& e). Following the development presented in the Methods section, our model includes the minimum features necessary to capture the salient features of spectral hole burning. 
The key predictions of this model are the decay rate $\Gamma_{\rm R}$ and frequency shift $\Delta \Omega_{\rm R}$ of the probe (as a function drive intensity) given by
\begin{align}
\label{decay}
\Gamma_{\rm R}(J_{\rm D})
 =\! - \frac{2\Omega}{Q_0}\!  \int_0^\infty\!\! \frac{dE}{2\pi\hbar} \frac{T_2S_z}{1+T_2^2(\Omega\! -\!E/\hbar)^2} 
 \\
 \label{freqShift}
\Delta \Omega_{\rm R}(J_{\rm D}) 
 = \!  \frac{\Omega}{Q_0}\!  \int_0^\infty\!\! \frac{dE}{2\pi\hbar} \frac{T^2_2(\Omega\! - \! E/\hbar)S_z}{1+T_2^2(\Omega\! -\!E/\hbar)^2} 
\end{align}
where the dimensionless parameter $Q_0^{-1}$ is given by $(\pi \mathscr{P} \gamma^2/\rho v^2)$, $\gamma$ is the deformation potential quantifying the coupling between TLS and phonons, $\rho$ is the material density, and $\mathscr{P}$ is the defect density of states (see Methods section) \cite{Golding1973,Golding1976b}. Here, $T^{-1}_2$, is the effective the TLS dephasing rate, and $S_z$ is the steady-state defect population inversion ($S_z \equiv N_e - N_g$ where $N_e$ and $N_g$ are the excited and ground state populations of the acoustic TLS.). When the detuning between the probe and drive is much greater than the TLS decay rate $T^{-1}_1$ ($|\Omega_{\rm D} - \Omega| \gg T_1^{-1}$) $S_z$ is given by 
\begin{equation}
\label{popInv}
S_z \approx - \frac{\tanh \left(E/2 k_B T\right) }{1+ [1+T_2^2(\Omega-E/\hbar)^2]^{-1} (J_{\rm pr}/J_{c,{\rm pr}})+[1+T_2^2(\Omega_{\rm D}-E/\hbar)^2]^{-1}(J_{\rm D}/J_{c,{\rm D}})}
\end{equation}
where $J_{\rm pr}$ is the acoustic intensity of the probe, and $J_{c,{\rm pr}}$ and $J_{c,{\rm D}}$ are critical intensities (described below). 

Equation \ref{decay} shows that the decay rate $\Gamma_{\rm R}$ is determined by the population inversion of the defect ensemble in a band of frequencies centered on $\Omega$. For low probe and drive powers the ensemble of acoustic atoms remains in thermal equilibrium at temperature $T$, i.e. $S_z = - \tanh(E/2 k_B T)$, and the acoustic TLSs behave like a linear absorber. In contrast, Eq. \ref{popInv} shows that an intense probe ($J_{\rm pr} \gg J_{c,{\rm pr}}$) forces $S_z \to 0$ near $\Omega$, saturating absorption by the acoustic atoms. 

In addition, an intense drive at a distinct frequency $\Omega_{\rm D}$ can saturate the dissipation of a weak probe. Large drive intensities send the ensemble of acoustic atoms into a nonequilibrium steady-state where $S_z$ approaches zero in a band of frequencies near $\Omega_{\rm D}$. Equation \ref{popInv} shows that the width of this band $2\pi \Delta \nu_{\rm SH}$, i.e. the phononic spectral hole, is given by 
\begin{equation}
\label{SHLW}
\Delta \nu_{\rm SH} = \frac{1}{\pi T_2}\left(1 + \sqrt{1+ J_{\rm D}/J_{c,{\rm D}}} \right) 
\end{equation}
when $\Omega \ {\rm and} \ \Omega_{\rm D} \gg T_2^{-1}$,
which is the same expression for the spectral hole width in strongly inhomogeneously broadened systems \cite{Siegman1986},
and shows that the drive saturates the dissipation of a weak probe when $(2\pi) \Delta \nu_{\rm SH} > 2|\Omega-\Omega_{\rm D}|$. 

Equations \ref{decay} and \ref{freqShift} show that phononic spectral hole burning spectroscopy provides a number insights and tools to manipulate the ensemble of acoustic atoms in a system. For sufficiently long $T_2$ (narrow sampling of $S_z$), measurements of $\Gamma_{\rm R}$ and $\Delta \Omega_{\rm R}$ determine the nonequilibrium steady-state population of the ensemble of the acoustic atoms. Additionally, $\Gamma_{\rm R}$ and $\Delta \Omega_{\rm R}$ depend sensitively on the dephasing time $T_2$, which is determined by the nature and strength of direct interactions between acoustic TLS, and quantifies the strength of spectral diffusion.

\begin{figure*}
  \includegraphics[width=\textwidth]{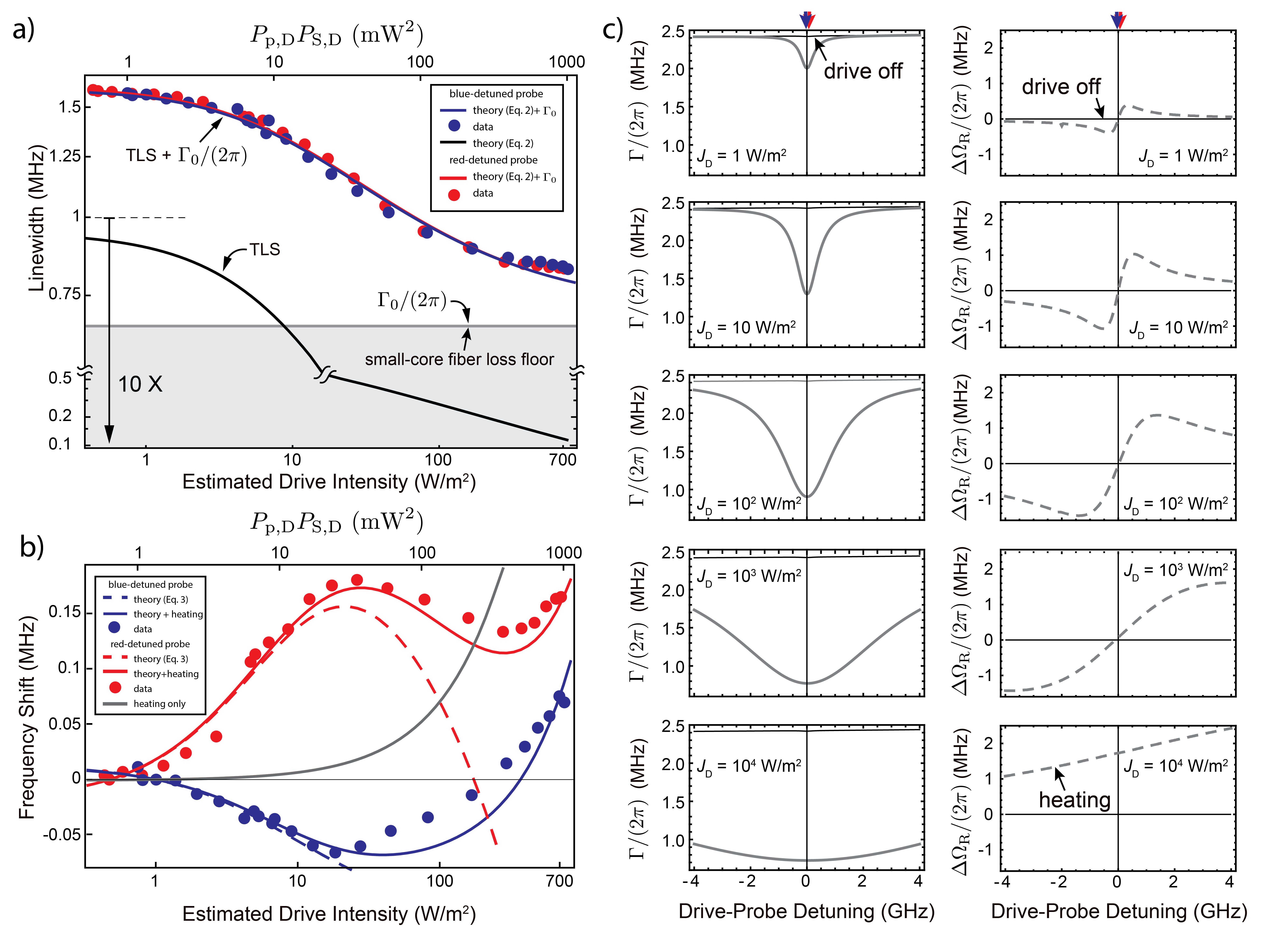}
  \caption{Theory and experiment for the acoustic probe a) linewidth and b) frequency shift as a function of drive intensity (and optical drive powers). The two probe-drive phonon detunings are $-65$ MHz for the red-detuned probe, and $+23$ MHz for the blue-detuned probe. The gray region in a) is the loss floor specific to UHNA-3 fiber, and the black line is the predicted absorption when the loss floor is set to zero. Phononic spectral hole burning reduces nonlinear absorption by $\sim$90\%. The effects of heating are negligible in a). c) Predicted linewidth and frequency shift as a function of probe-drive detuning and drive intensity. Acoustic TLS parameters of the text, $T \approx 1.15$ K (plus the effects of heating as described in the Supplemental Information D, E \& F), and the fitted value of $T_2 = 1.2$ ns are used to evaluate Eq. \ref{decay}. The probe intensity is set to zero. The red and blue arrows in c) indicate the detunings used in our experiment.}
\label{analysisResults}
\end{figure*}

{\it Theory experiment comparison:}
In this section we compare the phonon linewidth and frequency shift data extracted from the spectra of Fig. \ref{systemSketch-2} and displayed in Fig. \ref{analysisResults} to the theory of phononic spectral hole burning given in Eq. \ref{decay} and \ref{freqShift}. 
For this comparison, we model the total acoustic decay rate as $\Gamma = \Gamma_0+\Gamma_{\rm R}$, where $\Gamma_0$ quantifies the intensity-independent background losses specific to the fiber, and we obtain the frequency shift by subtracting the measured peak frequencies (black dots of Fig. \ref{systemSketch-2}d \& e) at the drive intensity of interest and the lowest drive intensity. 
Prior measurements of the acoustic dissipation rate characterize the ensemble of defects, yield the critical intensity, and establish the magnitude of the intensity-independent background dissipation $\Gamma_0$ in our optical fiber system \cite{Behunin2016,Kharel2016}. These measurements give $Q_0 \approx 1160$, $\Gamma_0 \approx (2\pi) 0.67$ MHz, and critical intensity $J_{c,{\rm D}} \approx J_{c,{\rm pr}} = 1.3$ W/m$^2$ (at 1.15 K). 

Simulations of the optical and acoustic modes relate the optical powers (used to generate phonons) to the intensity of the probe and drive. The intensity of the drive is given by $J_{\rm D} \approx (v \Omega_{\rm D}/\omega_{\s,{\rm D}} \Gamma A_{\rm eff}) G_B P_{\p,{\rm D}} P_{\s,{\rm D}}$ where $P_{\p,{\rm D}}$ and $P_{\s,{\rm D}}$ are the optical powers for the pump and Stokes fields that generate the drive, and $A_{\rm eff}$ ($1.6 \ \mu$m$^2$) is the mode field area of the acoustic drive. We estimate a dephasing time for defects in this system of {$T_2 \approx 1.3$ ns}, which is derived from empirically obtained properties of the defect ensemble and the theory of acoustic dissipation in glass \cite{Behunin2016}. Using this analysis we can calculate the effects of phononic spectral hole burning using Eq. \ref{decay} \& \ref{freqShift} (see Supplemental Information Sec. B for details).

We compare our measurements with the prediction for the frequency shift and linewidth (i.e. $\Gamma/2\pi$) based on the parameters above in Fig. \ref{analysisResults}. Equation \ref{decay} provides a good description of the linewidth, and Eq. \ref{freqShift} explains the measured frequency shift for low intensities. We find that Eq. \ref{freqShift} fails to explain our observations for intensities in excess of 20 W/m$^2$ when the measured temperature of the liquid helium bath is used to compute the frequency shift (dashed lines Fig. \ref{analysisResults}b). We attribute this discrepancy to heating in the fiber core that leads to an additional frequency shift above that given by Eq. \ref{freqShift}. We estimate the effect of heating in the fiber core by using the intrinsic properties of TLS \cite{Phillips1987} (see Supplemental Information Sec. D).  

Figure \ref{analysisResults} displays a more refined prediction of the linewidth and frequency shift, including the effects of heating, as solid blue and red lines. We obtain the displayed theory by optimizing the value of $T_2$, estimating the actual temperature in the fiber core, and using the theoretical prediction for defect dynamics in the fiber (see Supplemental Information Sec. D). The fitted value of $T_2 = 1.2$ ns agrees remarkably well with the estimate derived from the critical intensity obtained from other measurements \cite{Behunin2016}. 

The fitted value of $T_2$ and the theory-experiment agreement reveal details of the direct interaction between acoustic atoms. The phenomenological dephasing rate is given by $T_2^{-1} = (2 T_1)^{-1} + T^{-1}_{SD}$ where $T^{-1}_{SD}$ is the dephasing rate due to spectral diffusion, and recall that $T_1$ is the acoustic atom's excited state lifetime. Spectral diffusion occurs when the energy of a TLS is perturbed by thermally-induced transitions of its neighbors. Such perturbations are mediated by direct interactions between acoustic atoms. Since the excited state lifetime of the relevant TLS in our system ($T_1 \approx 79$ ns \cite{Behunin2016}) is much greater than $T_2$, we conclude that spectral diffusion dominates TLS dephasing. This result shows the importance of direct interactions between acoustic atoms and explains why cross relaxation produces such a large spectral hole in this system.   

The data clearly demonstrate that an intense acoustic drive extends probe phonon lifetimes and modifies the sound speed. 
We attribute the dissipation floor in this system, gray region of Fig. \ref{analysisResults}a, to scattering losses associated with inhomogeneities in the fiber geometry. The relatively small difference in the saturation of defect absorption within a bandwidth of 65 MHz from the probe phonon frequency at 9194 MHz, suggests a hard 130 MHz lower limit on the width of the phononic spectral hole burnt by the drive beam. This limit is consistent with Eq. \ref{SHLW} which gives a lower 490 MHz limit to the spectral hole size based on the fitted value of $T_2$. 

\begin{figure}
\begin{center}
\includegraphics[width=0.4\textwidth]{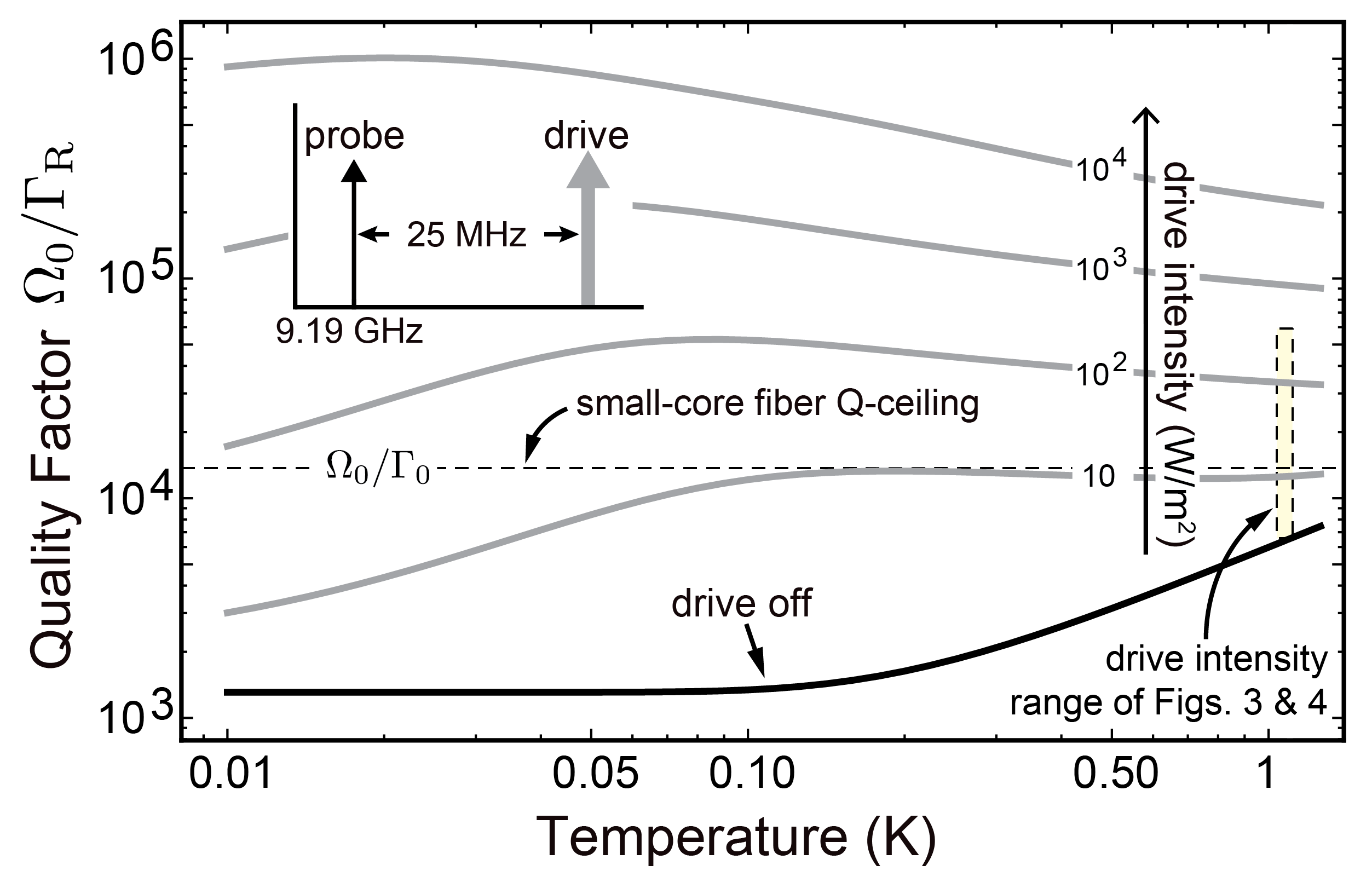}
\caption{Acoustic quality factor with temperature and drive intensity in a system where absorption by acoustic atoms dominates dissipation $(\Gamma_{\rm R} \gg \Gamma_0)$. The respective probe and drive frequencies are $9.19$ GHz and $9.44$ GHz (-$25$ MHz probe-drive detuning). We use the tunneling state model and the measured properties of acoustic atoms in our system \cite{Behunin2016} to determine the temperature dependence of $T_1$ and $T_2$ (see supplemental information). Each gray curve represents the quality factor for a different drive intensity. The vertical dashed yellow box at $T = 1.15$ K corresponds to the temperature and intensity range of our measurements.}
\label{results-2}
\end{center}
\end{figure}

We display theoretical predictions of the acoustic absorption and frequency shift in Fig. \ref{analysisResults}c and Fig. \ref{results-2} to explore the effects of hole burning for probe-drive detunings and temperatures beyond the capabilities of our apparatus (optical drive frequencies outside the C-band and temperatures below 1 K). Figure \ref{analysisResults}c shows that spectral hole widths well in excess of 7 GHz and that large tunable frequency shifts in excess of the linewidth may be possible near the highest (acoustic) intensities achieved in our experiments ($1000$ W/m$^2$ or acoustic powers of 1.6 nW). Figure \ref{results-2} shows the temperature dependent quality factor at a variety of drive intensities for systems with acoustic atom dominated acoustic dissipation (i.e. $\Gamma_{\rm R} \gg \Gamma_0$). These results show that phononic spectral hole burning may yield acoustic quality factors in excess of $10^6$ for low drive powers ($\sim$16 nW) when background losses are small. 

In conclusion, we have demonstrated that continuous acoustic driving can greatly reduce intrinsic forms of acoustic dissipation by more than 90\% over a wide band of phonon frequencies through nonlinear saturation (i.e., spectral hole burning) in silica. 
We showed that direct interactions among acoustic TLS enable efficient cross relaxation processes that produce a transparency window covering a wide fractional bandwidth ($\sim$100\%) at modest (nW) drive powers.
We have developed a simple model that explains the dispersive and dissipative changes that we observed with nonlinear phonon saturation spectroscopy. This model, which builds on established phenomenological models of TLS based glass physics, explains the nonlinear measurements in a self-consistent manner, and only requires TLS parameters obtained from independent experiments. 

The ability to turn glass into a highly transparent phononic material with an external drive field can be used to dramatically enhance the performance of a number of phononic systems in amorphous media. To appreciate the potential for improved performance, it is important to note that intrinsic dissipation of glass makes up a small component of the overall dissipation rate we observe at high drive intensities ($(2\pi)$ 0.8 MHz). The majority of this loss floor likely originates from the high concentration of dopants in our fiber system \cite{Behunin2016,Dragic2009}. Without these extrinsic sources of dissipation (scattering and leakage), the transient pulse saturation measurements performed by Golding {\it et al.} suggest a path to radically reduced overall dissipation rates. Despite dissipation limits imposed by diffraction losses \cite{Golding1973}, Golding {\it et al.} measured coherence times in bulk silica (dopant free) translating to Q-factors of $0.3 - 1 \times 10^6$ \cite{Golding1973,Golding1976b} at $20-100$ milli-Kelvin temperatures. These observations are consistent with the quality factor predictions given in Fig. \ref{results-2}, which apply to systems with small amounts extrinsic dissipation.  

This comparison suggests that dramatically enhanced phonon coherence may be achieved in amorphous systems when extrinsic sources of loss (e.g. dopants and diffraction) are eliminated. Silica microspheres \cite{Park2009,Bahl2011,Bahl2011a}, microtoroids \cite{Anetsberger2009,Schliesser2008,Schliesser2009,Riviere2011}, and wedge resonators \cite{Lee2012} are compelling candidates to test these ideas since whispering gallery mode phonons interact with atomically smooth surfaces, reducing extrinsic sources of scattering and tunneling losses. Putting all of this together, phononic spectral hole burning may offer a path towards ultra-high coherence phonon modes to match the already exceptional optical properties of silica based photonics.

We have shown how external fields control the elastic properties of a system, opening the door to on-demand phononic switching and the use of glasses as low loss phononic media. For instance, these tools allow the resonance condition of a cavity, the guidance properties of a waveguide, or optomechanical couplings to be engineered {\it in situ}. Low losses lead to long phonon lifetimes that enhance information processing capabilities such as optical pulse delay \cite{Okawachi2005} and the memory of stored optical information \cite{Fiore2011}. Enhanced Brillouin gain, brought about by suppressed dissipation, may enable ultra-low threshold lasing \cite{Stokes1982,Lee2012}, broad-band frequency comb generation \cite{Kang2009,DelHaye2007,Kippenberg2011}, and new regimes of slow light \cite{Boyd2009} and optomechanical (or Brillouin scattering) induced-transparency \cite{Weis2010,Kim2015}. At high intensities and low temperatures phonon lasing \cite{Vahala2009,Mahboob2013} may be possible in microspheres, microtoroids, or microdisks.
Although we have focussed on amorphous materials in this paper, the tools we have presented may enable unprecedented levels of performance in an array of cutting edge-device technologies.  

\section{Acknowledgments}
RB and PR would like to thank Paul Fleury for a number of thoughtful criticisms and suggestions. 

\section{Methods}
\subsection{Tunneling state physics \& theory of phononic spectral hole burning}
\begin{figure}
\begin{center}
\includegraphics[width=0.5\textwidth]{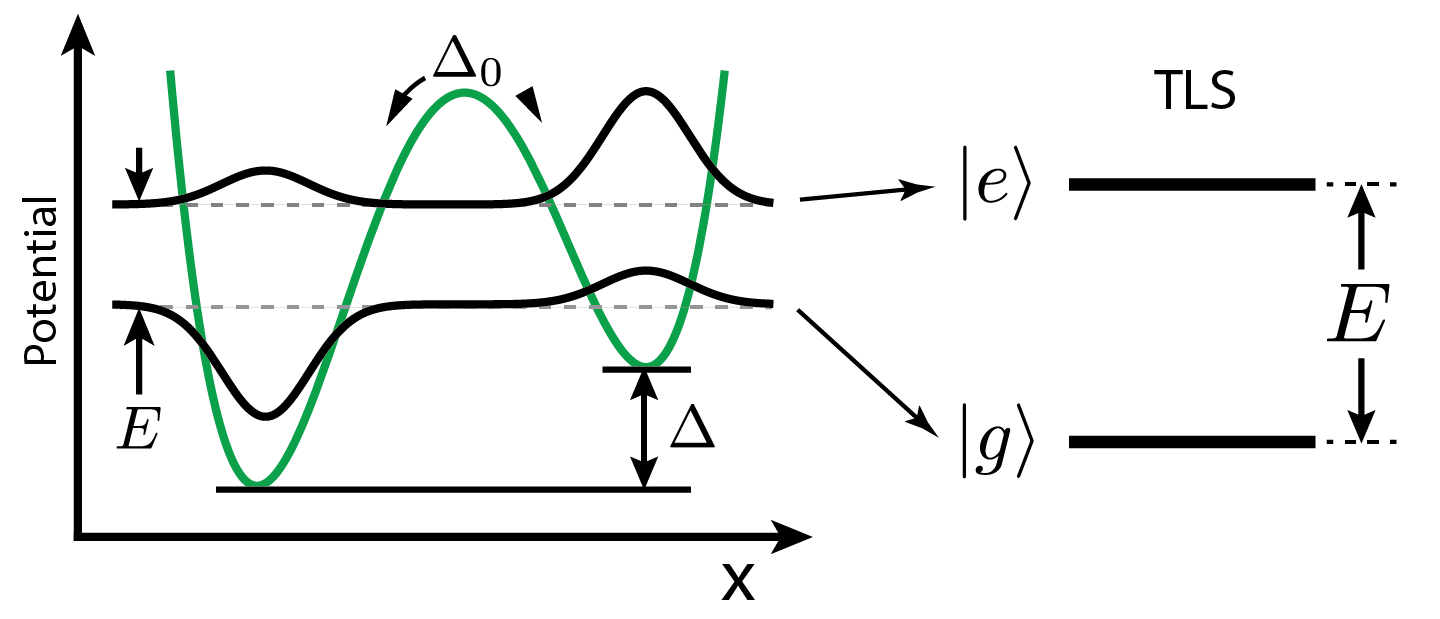}
\caption{Defect double-well potential of asymmetry $\Delta$ and tunneling strength $\Delta_0$. Excited $\left| e \right>$ and ground $\left| g \right>$ eigenstates are gapped by energy $E$.}
\label{TLS-diagram}
\end{center}
\end{figure}

Tunneling state defects are hypothesized to originate from atoms (or collections of atoms) residing in double-well potentials of asymmetry $\Delta$ and tunneling energy $\Delta_0 = \hbar \omega_c e^{-\lambda}$ (Fig. \ref{TLS-diagram}) where $\omega_c$ is the natural frequency of each well, and $\lambda = d\sqrt{2mV_0/\hbar^2}$ ($V_0$ is the barrier height, $d$ is the `separation' between the potential's two minima, and $m$ is the tunneling atom(s) mass) quantifies the wavefunction overlap between the two wells \cite{Phillips1987}. At low temperatures these tunneling states can be modeled as effective two-level systems (TLS) of energy $E = \sqrt{\Delta^2+\Delta_0^2}$  (Fig. \ref{TLS-diagram}) \cite{Anderson1972,Phillips1987}. 

Strain-induced perturbations of the TLS double-well potential couple defects and phonons. The coupling leading to phononic spectral hole burning is described by the interaction Hamiltonian
\begin{equation}
\label{Hamiltonian}
H_{\rm int} = \sum_{j,q} \  \hbar g_{q,j}\sigma^+_{j}   b_q + {\rm H.c.}
\end{equation}
where the indices $j$ and $q$ respectively label defects and the phonon modes of the system, $\sigma^+_{j}$ is the raising operator for the TLS, $g_{q,j}$ is the defect-phonon coupling rate (described in detail below) \cite{Behunin2016b}, $b_q$ is the annihilation operator for the $q$th phonon mode, and the rotating wave approximation (RWA) has been used. 
TLS also couple non-resonantly to phonons, to other defects (leading to spectral diffusion), and to the electromagnetic field \cite{Phillips1987}. Of these additional processes only spectral diffusion is relevant to our analysis, and we account for its effects below in a phenomenological manner. 

Large systems are assumed to contain an ensemble of TLS having a distribution $f(\Delta,\Delta_0)$ of asymmetries, tunneling strengths, orientations, and positions. We assume that processes involving a large number of defects are described statistically using $f(\Delta,\Delta_0)$. For ensemble averaging, we use the distribution posited by the standard tunneling state model $f(\Delta,\Delta_0) = \mathscr{P}/\Delta_0$, where $\mathscr{P}$ is a constant with units of number density per unit energy, that assumes a uniform distribution of defect positions and orientations \cite{Anderson1972,Phillips1987}. 

Phononic spectral hole burning is characterized by the modification of the steady-state dynamics of the probe phonon of classical amplitude $b_{\rm pr}$ and frequency $\Omega$ by a drive of classical amplitude $b_{\rm D}$ and frequency $\Omega_{\rm D}$ driven strongly on resonance. For a system driven at two frequencies, the inherent nonlinearity of the TLS response generally results in the generation of a large number of sidebands. However, in the limit when $|b_{\rm D}| \gg |b_{\rm pr}|$ and when the detuning between the drive and the probe is much greater than the TLS decay rate $T_1^{-1}$, i.e. $|\Omega_{\rm D} - \Omega| \gg T_1^{-1}$, the contribution from higher order sidebands can be dropped. 
In this limit the physics is described by the coupled Bloch equations given by 
\begin{align}
\label{B-Eqs}
   \dot{S}_{z} & = -\frac{1}{T_1}(S_z - w_{eq}(E)) \\
    & \quad \quad \quad  - i 2 \left((\gp b_{\rm pr} + \ga b_{\rm D}) S^+ - H.c. \right)  \nonumber \\
   \dot{S}^+   &   = \left( \frac{i}{\hbar}E - \frac{1}{T_2} \right) S^+ 
    -  i(\gp^* b_{\rm pr}^\dag + \ga^* b_{\rm D}^\dag) S_z   \\
   \dot{b}_{\rm pr}    &   = - i \Omega b_{\rm pr} - i \sum_j  \gp S^{+*}
 \end{align} 
where $S_z$ and $S^+$ are the mean values of $\sigma_z$ and $\sigma^+$ respectively, $\sigma_z$ is the $z$th Pauli matrix, $w_{eq}(E) = -\tanh(E/2 k_B T)$ is the thermal expectation value of $\sigma_z$ for temperature $T$, and the defect label $j$ has been suppressed in coupling rates and TLS operators. Above, the quantum and thermal effects of the phonon field and neighboring defects is accounted for in the defect decay $T^{-1}_1$ and the dephasing $T_2^{-1}$ rates. $T^{-1}_1$ characterizes the rate at which an excited defect decays through the emission of a phonon, and $T^{-1}_2$ phenomenologically characterizes the dephasing rate given by $T_2^{-1} = T_1^{-1}/2 + T_{\rm SD}^{-1}$ where $T_{\rm SD}$ quantifies the timescale for spectral diffusion \cite{Phillips1987,Behunin2016b}. 
 
The effect of phononic spectral hole burning is derived in two steps. First, the steady-state dynamics of the defects, driven by the probe and drive phonons, is obtained. This formal solution, depending on the drive phonon field, is then used to find the effects on the probe phonon dynamics.

Both the probe and drive phonons drive oscillations of $S^+$. If the detuning $\Omega-\Omega_{\rm D}$ is much greater than $T_1^{-1}$ then $S_z$ is approximately time-independent and the steady-state solution for $S^+$ is given by  
\begin{equation}
\label{S+}
  S^+ \approx  -  \frac{i \gp^* b_{\rm pr}^\dag S_z}{ i \left(\Omega - \frac{1}{\hbar}E \right) + \frac{1}{T_2}} 
   -  \frac{i\ga^* b_{\rm D}^\dag S_z}{ i \left(\Omega_{\rm D} - \frac{1}{\hbar}E \right) + \frac{1}{T_2}}. 
\end{equation} 
The steady-state value for $S_z$, the defect population inversion, is obtained by plugging Eq. \ref{S+} into  Eq. \ref{B-Eqs} giving Eq. \ref{popInv}.

With the solution for $S^+$ and $S_z$ the defect and drive-beam influenced dynamics of the probe phonon mode is obtained by plugging the solution for $S^+$ into the equation of motion for $b_{\rm pr}$ resulting in 
$- i \sum_j  \gp S^{+*} \to (-i \Delta \Omega_{\rm R} - \Gamma_{\rm R}/2) b_{\rm pr} +...$
where $\Gamma_{\rm R}$ is the decay rate characterizing resonant acoustic absorption by the defects and $\Delta\Omega_{\rm R}$ characterizes the concomitant shift in the phonon's frequency.   For the case where a large ensemble of defects interact with plane acoustic waves the frequency shift and decay rate are given by Eq. \ref{decay} where the critical intensity is given by
\begin{equation}
\label{criticalIntrensity}
J_c = \frac{ \hbar^2 \rho v^3}{2 \gamma^2 T_1 T_2}
\end{equation}
and where the phonon intensity is given by $J = \hbar \Omega v b^\dag b/(L A_{eff})$. 

\section{Supplemental Information}

\subsection{Cross relaxation processes in disordered media}
\begin{figure}
\begin{center}
\includegraphics[width=0.75\textwidth]{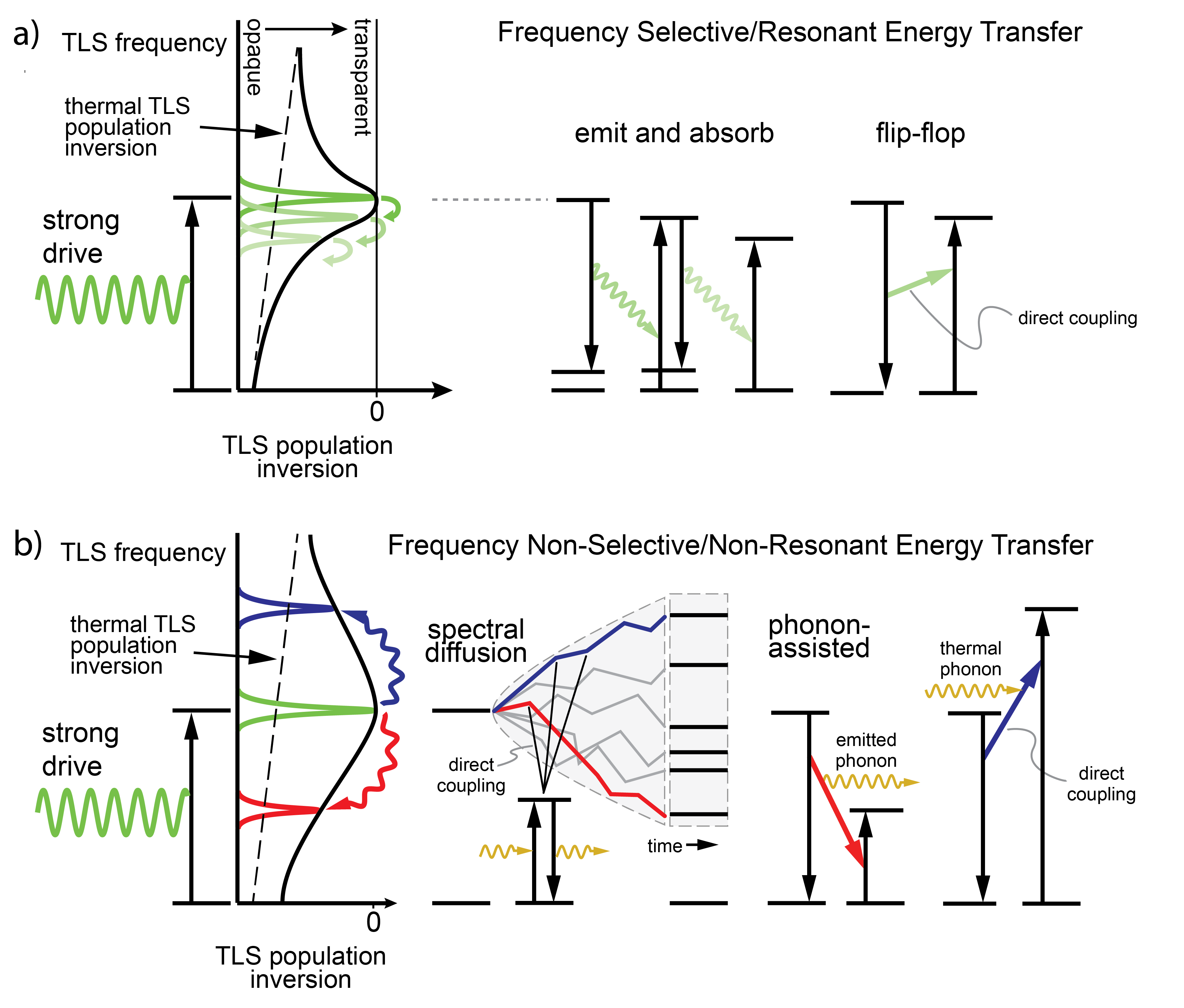}
\caption{a) Resonant and b) non-resonant energy transfer processes in a disordered material. Strong driving in conjunction with energy transfer renders disordered media transparent in a wide spectral hole.}
\label{energyTransfer}
\end{center}
\end{figure}

Taken at face value, the standard tunneling state model of low-temperature glass implies
that a number of complex processes contribute to phononic spectral hole burning (Fig. \ref{energyTransfer}) \cite{Phillips1987,Behunin2016}. 

At its heart, hole burning is accomplished by populating the excited states of acoustic TLS with a strong drive, rendering these TLS transparent. Subsequently, this energy is transferred throughout the system by an array of a) frequency selective and b) frequency non-selective cross-relaxation processes \cite{Siegman1986} illustrated in Fig. \ref{energyTransfer}. 

For example, excited acoustic atoms decay into a band of frequencies (linewidth) through spontaneous emission. In contrast with atoms, TLS in glass have no innate energy scale and have energies distributed over a wide band. Hence, spontaneously emitted phonons can be reabsorbed, and reemitted over and over resulting in a spectral hole width only limited by the input power (Fig. \ref{energyTransfer}a). 

In addition, direct interactions among acoustic atoms mediate `flip-flop' processes (Fig. \ref{energyTransfer}a), as well as a number of non-resonant (phonon assisted) energy transfer channels \ref{energyTransfer}b). In spectral diffusion the width of a narrow band of TLS energy levels grows in time (diffuses) as neighboring TLS undergo spin-flips induced by emission and absorption of thermal phonons. In this process the energy of TLSs excited by the drive diffuses over a large band. In addition, resonant absorption or emission of phonons combined with a flip-flop can transfer energy between TLS of distinct energies. These processes distribute energy over a wide spectral window and lead to the large spectral hole bandwidth observed in our experiments \ref{energyTransfer}b). 

\subsection{Brillouin gain and acoustic mode area}
In this section we discuss the nonlinear susceptibility (Brillouin gain) that enables our laser based form of nonlinear phonon spectroscopy. The Brillouin gain coefficient given by
\begin{equation}
\label{Gain}
G_B  = \frac{n^7 p_{12}^2 \omega_{\s}^2}{\rho v c^3 \Gamma} \frac{1}{A} 
\end{equation}
where $n$ is the effective index of refraction for the optical fields, $p_{12}$ ($p_{12} = 0.12$ in silica) is the photoelastic constant, $\rho$ is the material density, $v$ and $c$ are the sound and light speeds, $\Gamma$ is the phonon decay rate, and $A$ is the effective overlap area between the optical and acoustic fields \cite{Boyd2003,Rakich2012}. 

\begin{figure}
\begin{center}
\includegraphics[width=0.4\textwidth]{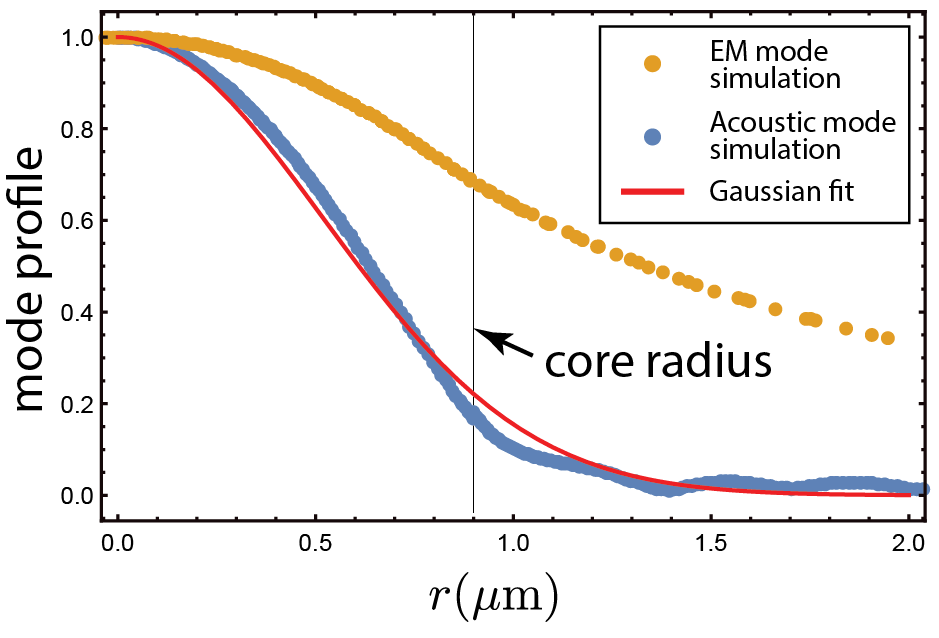}
\caption{Electromagnetic and acoustic mode profiles as a function distance from the fiber center.}
\label{modeProfiles}
\end{center}
\end{figure}

The effective area is given by 
\begin{equation}
\label{ }
A =  \bigg| n^2 \sqrt{\rho} L^{3/2} \int d^2 x \ \mathcal{E}_{\p}\mathcal{E}^*_{\s} u_q^* \bigg|^{-2}
\end{equation}
where a small transverse and longitudinal component of the respective acoustic and electric eigenfunctions $\{u_q, \mathcal{E}_{\p},\mathcal{E}_{\s}\}$ has been neglected. The acoustic field satisfies the time harmonic elastic equations of motion (see defect-phonon coupling section) and is normalized such that $\int d^3 x \ \rho |{\bf u}_q|^2 =1$. The electric field eigenfunctions satisfy the time harmonic Maxwell's equations in the fiber and are normalized as $\int d^3 x\ \varepsilon |\mathcal{E}|^2 = 1$ where $\varepsilon$ is the local value of the permittivity. The effective index, modal sound speed, and electric and acoustic field eigenfunctions (see Fig. \ref{modeProfiles}) are found with finite element simulations using the material properties of the fiber core and cladding. The density, index of refraction, and sound speed in the (Nufern UHNA-3) fiber core and cladding are given respectively by; $\rho = 2666$ kg/m$^3$, $n_{\rm core} = 1.485$ and $v_{\rm core} = 4740$ m/s (we estimate the shear speed in the core as $3092$ m/s) for the core \cite{Dragic2009}, and $\rho = 2202$ kg/m$^3$, $n_{\rm clad} = 1.444$, $v_{\ell,{\rm clad}} = 5944$ m/s (longitudinal) and $v_{t,{\rm clad}} = 3764$ m/s (transverse) \cite{Jen1986}. With these parameters numerical evaluation of the acousto-optic mode overlap gives $A \approx 0.09 A_{\rm core}$, where the core cross section is given by $A_{\rm core} = \pi (0.9)^2 \ \mu$m$^2$, the effective index $n \approx 1.449$, and the modal sound speed $v = 4910$ m/s, giving $G_B \approx 0.63 \ (51\ {\rm MHz}/\Delta \nu)$ (Wm)$^{-1}$ where $\Delta \nu$ is the phonon linewidth ($\Delta \nu = 51$ MHz at room temperature). 

The effective phonon mode area, relevant to estimating the acoustic intensity generated via SBS, is approximately  
$A_{\rm eff} \approx (\int d^2 x |{\bf u}_q(r)|^2 )^2/ \int d^2 x |{\bf u}_q(r)|^4)$ where $|{\bf u}_q(r)| \approx \exp\{- \beta (r/r_c)^2\}$, where $r_c = 0.9 \mu$m is the fiber core radius and $\beta \approx 1.5$, is the phonon mode profile shown in Fig. \ref{modeProfiles}.  
Simulations of the acoustic field give $A_{\rm eff} \approx 1.6 \ \mu$m$^2$.

\subsection{Experimental setup}

\begin{figure*}[t!]
\begin{center}
\includegraphics[width=\textwidth]{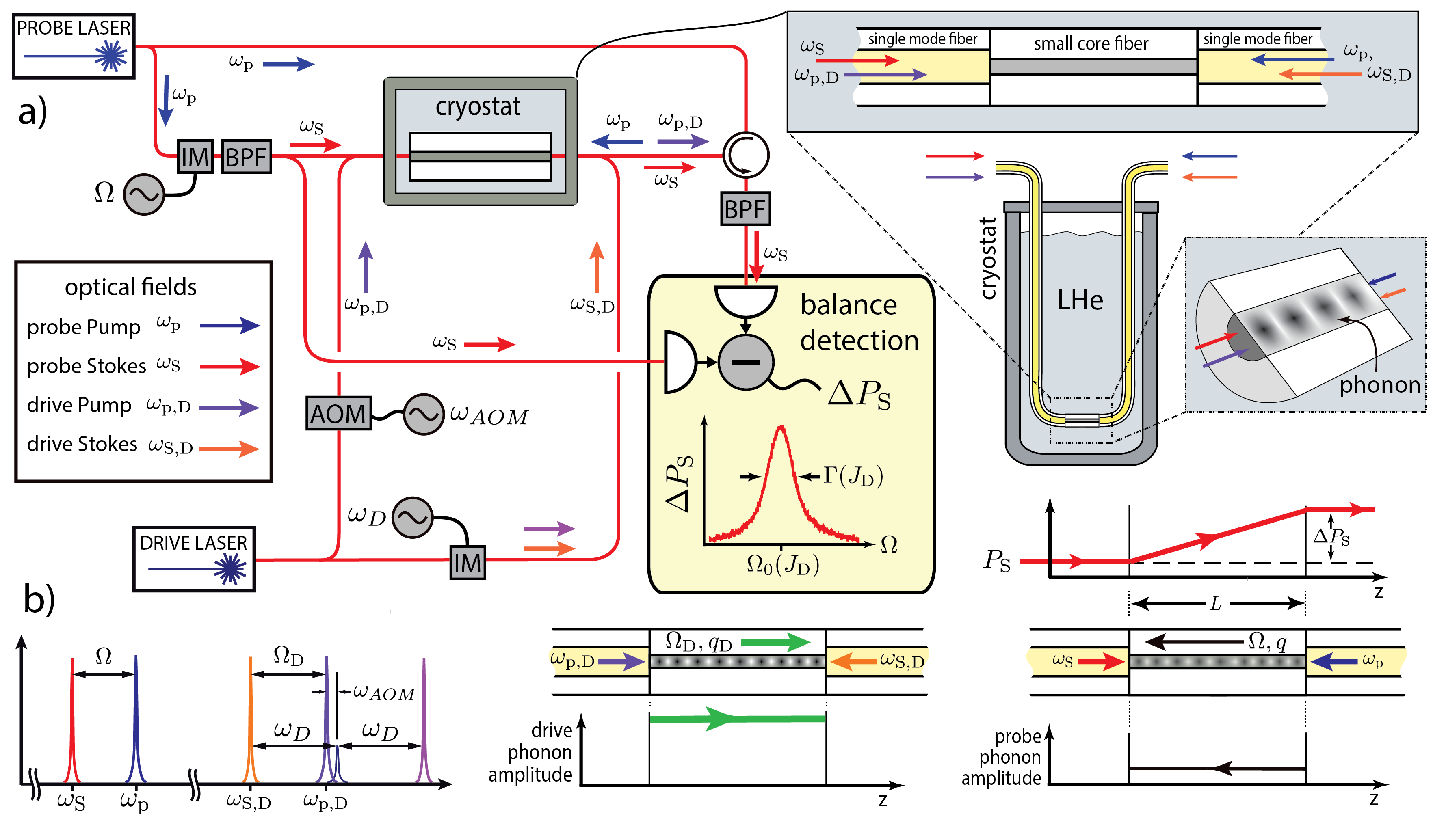}
\caption{a) Apparatus used to measure phonon spectral hole burning. The probe and drive Stokes beams are synthesized with an intensity modulator (IM). A BPF removes the undesired high frequency sideband for the probe. Brillouin gain spectra of the probe Stokes beam is measured using balanced detection. b) Optical frequencies used to generate drive and probe phonons.}
\label{holeBurningSetup}
\end{center}
\end{figure*}

In this section we discuss the experimental setup (Fig. \ref{holeBurningSetup}) in detail. 
First, the probe laser source (distributed feedback laser) is split in to two arms; one to act as the probe pump beam, and the second arm to act as the probe Stokes beam (Fig. \ref{holeBurningSetup}a). The Stokes arm is synthesized with an intensity modulator and a bandpass filter (BPF). The probe Stokes beam is split into a signal arm, which counter-propagates with the probe pump beam through the sample and is amplified via SBS, and a reference arm. The net amplification of the Stokes beam is measured by comparing the signal and reference arms on a balance detector (BD). At various points in each of these arms; the optical power is amplified (using erbium doped fiber amplifiers) or  attenuated (using variable optical attenuators), and the  polarization is controlled with polarization controllers. 

The drive beam is synthesized, using a tunable external cavity laser, and controlled in the same manner as the probe beam. However, in order to generate multiple drive phonon frequencies no filtering is used after the IM, and hence both sidebands and the residual carrier for the drive Stokes beam pass through the sample. Both of these drive Stokes arm sidebands interact with the drive pump through Stokes and anti-Stokes scattering processes. We break this symmetry between the two sidebands by shifting the drive pump line $39.5$ MHz (much greater than the $\sim 2$ MHz coupling bandwidth for SBS) by an acousto-optic modulator (AOM), allowing only one drive Stokes arm sideband to interact with the drive pump at a time (Fig. \ref{holeBurningSetup}b). 

The distinct elastic properties of the fiber leads and the fiber sample ensures that the phonons are only generated in the sample under study. This form of multiplexing (Brillouin multiplexing) is accomplished because all the optical fiber used in the setup, except for the sample under test, is SMF-28 with a Brillouin resonance near $11$ GHz and well separated from the Brillouin resonance of the sample near $9$ GHz. 

Optical multiplexing isolates the desired signal contained in probe light. We place a polarizer and a BPF before the signal arm input on the balanced detector. The filter passes probe frequencies and blocks the drive. The polarization of the probe beam is adjusted to optimize the signal, and the polarization of drive beam is adjusted to maximize phononic spectral hole burning and to prevent drive power from reaching the detector. 

\subsection{Temperature Calibration}
\begin{figure}
\begin{center}
\includegraphics[width=0.75\textwidth]{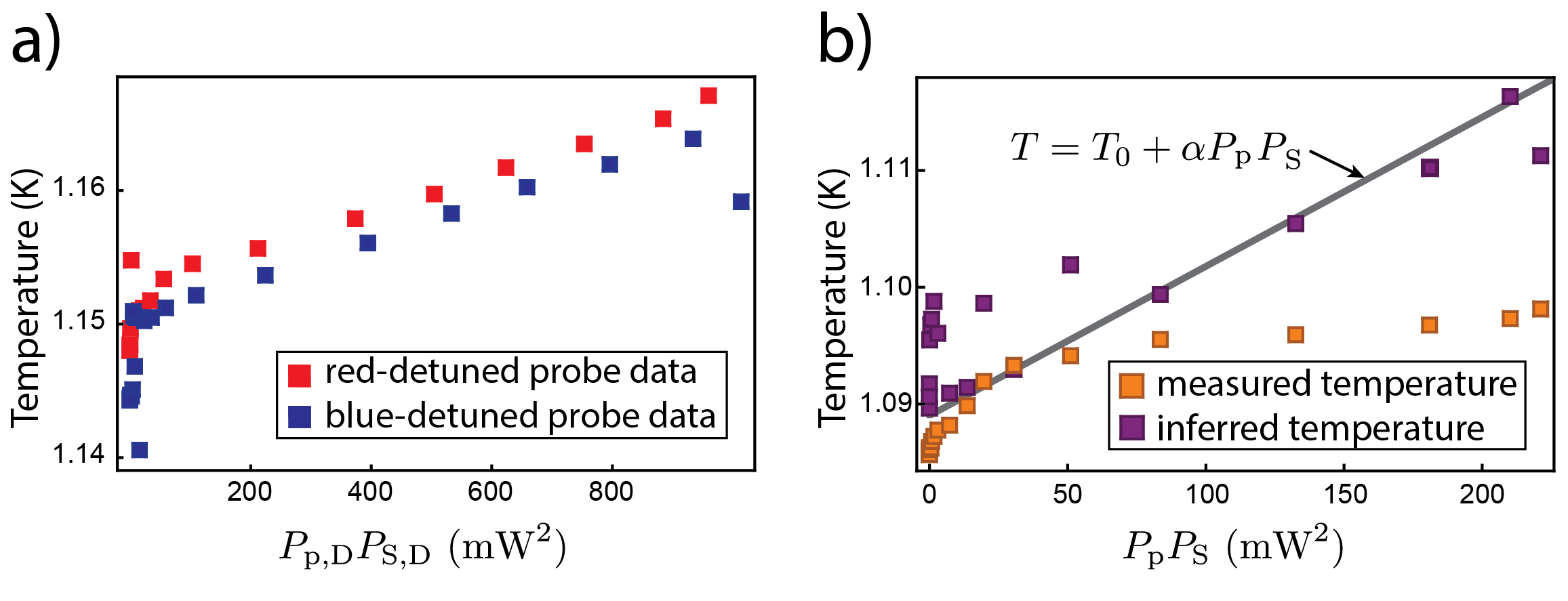}
\caption{Temperature calibration data}
\label{TemperatureCal}
\end{center}
\end{figure}

In this section we demonstrate that heating occurs in the sample, and we 
estimate the temperature inside the fiber core.  
Heating in the fiber is established in Figure \ref{TemperatureCal} a which clearly shows that the measured temperature increases with the optical power used to synthesize the acoustic drive.

Heating leads to shifts in the probe phonon frequency, and thus must be accounted for in addition to frequency shifts due to spectral hole burning to explain the data of Fig. \ref{analysisResults}. In the absence of a drive, the tunneling state model predicts a temperature-dependent shift in the speed of sound given by 
$v(T)-v(T') \approx (v/\pi Q_0)\ln (T/T')$ \cite{Phillips1987}.
This shift in $v$ also leads to a shift in the peak frequency $\Omega_0$ of the probe excited through SBS given by
\begin{equation}
\label{DeltaFT}
\Omega_0(T) - \Omega_0(T') \equiv \delta \Omega_0(T,T')  \approx \frac{\Omega_0}{\pi Q_0}\ln (T/T')
\end{equation} 
if we assume that the refractive index does not change with temperature. 
Therefore, the frequency shift due to heating can be estimated by using Eq. \ref{DeltaFT} and the measured temperature. However, this analysis does not explain the difference between the predictions of our spectral hole burning model and the data of Fig. \ref{analysisResults}. We attribute this disagreement to a discrepancy between the measured temperature and the actual temperature in the fiber core.  

Equation \ref{DeltaFT} provides a tool to estimate the calibration error between the measured temperature and the actual temperature inside the fiber. In the absence of a drive (i.e. no spectral hole burning) we performed simultaneous and independent measurements of probe frequency $\Omega_0$ and temperature as a function of the optical power.  
Hence, the measured frequency used with Eq. \ref{DeltaFT} can be used to estimate the temperature in the core ${T_{\rm core} \approx T_0 \exp(\pi Q_0 \delta \Omega_0(T,T_0)/\Omega_0)}$ where we set the arbitrary reference temperature $T_0$ to the lowest measured temperature. The measured temperature and the temperature inferred from the frequency shift data is displayed in Fig. \ref{TemperatureCal} b), revealing that the temperature in the fiber core is likely several mK higher than the thermometer readout. 

The results of this analysis support a simple linear relationship for the temperature in the fiber core as a function of large optical power given by $T_{\rm core} = T_0 + \alpha P_\p P_\s$ where $\alpha$ is a fit parameter. We obtain the parameter $\alpha  \approx 129 \times 10^{-6}$ K (mW)$^{-2} $ by fitting Eqs. \ref{decay} and \ref{DeltaFT}, with $T \to T_0 + \alpha P_{\p,{\rm D}} P_{\s,{\rm D}}$, to the frequency shift and linewidth data of Fig. \ref{analysisResults}. As a consistency check the power dependence of the inferred temperature in the fiber is shown as the gray line of Fig. \ref{TemperatureCal} b) revealing that the fitted value (obtained from our hole burning data) explains the inferred temperature of the system in the absence of hole burning. 

\subsection{TLS parameters}

The dephasing rate $T_2 \approx 1.3$ ns (at 1.15 K) is obtained from the theoretical expression for the critical intensity (Eq. \ref{criticalIntrensity}) and $T_{1}$ \cite{Phillips1987} ($\gamma = \ 0.5$ eV), the measured value of the critical intensity \cite{Behunin2016}, and the material properties of germanium doped glass \cite{Dragic2009}. The product $\gamma^2 T_{1}$ appearing in the expression for the critical intensity is insensitive to the precise value of the deformation potential, hence this estimation for $T_2$ is remarkably insensitive to uncertainties in the details of the defect-phonon coupling in our system. 
\subsection{Acoustic atom decay and dephasing rates}     
In this section we discuss how the decay and dephasing rates of acoustic atoms were treated in our analysis.                                                                                                                                                                                                      For a TLS of energy $E$ and tunneling strength $\Delta_0$ in a bulk system the decay rate given by
\begin{equation}
\label{ }
\frac{1}{T_1} = \sum_\eta \frac{\gamma_\eta^2}{v_\eta^5} 
\frac{E \Delta_0^2}{2\pi \hbar^4 \rho} \coth \left( \frac{E}{2 k_B T} \right) 
\end{equation}
follows from Fermi's Golden rule and Eq. \ref{Hamiltonian}
where $\eta$ denotes the phonon polarization \cite{Phillips1987}. Here, $\gamma_\eta$ and $v_\eta$ are the deformation potential and sound velocity for $\eta$-polarized phonons. The minimum TLS lifetime $T_{1,{\rm min}}$, used to compute the critical intensity, is given by $T_1$ when $\Delta_0 = E$.  We assume that the deformation potential for shear phonons $\gamma_t$ is given by $\gamma^2_t = \gamma^2/2$ where $\gamma$ is the deformation potential for longitudinal phonons \cite{Golding1976b}. This shows $T_1 \propto \gamma^{-2}$, making $\gamma^2 T_1$ insenstive to the deformation potential as stated in the Sec. E. 

The dephasing rate is determined by $T_1$ and by spectral diffusion \cite{Phillips1987,Black1977,Behunin2016b}. For our system $T_2 \ll T_1$ which implies spectral diffusion dominated defect dephasing. A simple scaling argument gives spectral diffusion dephasing rate as $T^{-1}_{\rm SD} \propto P(k_B T) T$ in bulk systems. We use this scaling and the recently measured energy dependence of the defect density of states $P(E) \propto E^{0.28}$ \cite{Skacel2015}, to estimate the temperature dependence of defect dephasing. 

\subsection{Evidence for TLS-induced phonon dissipation in small core fiber at cryogenic temperatures}
In this section we outline the evidence for the presence of TLS-induced dissipation in our system. 

\begin{figure}
\begin{center}
\includegraphics[width=0.5\textwidth]{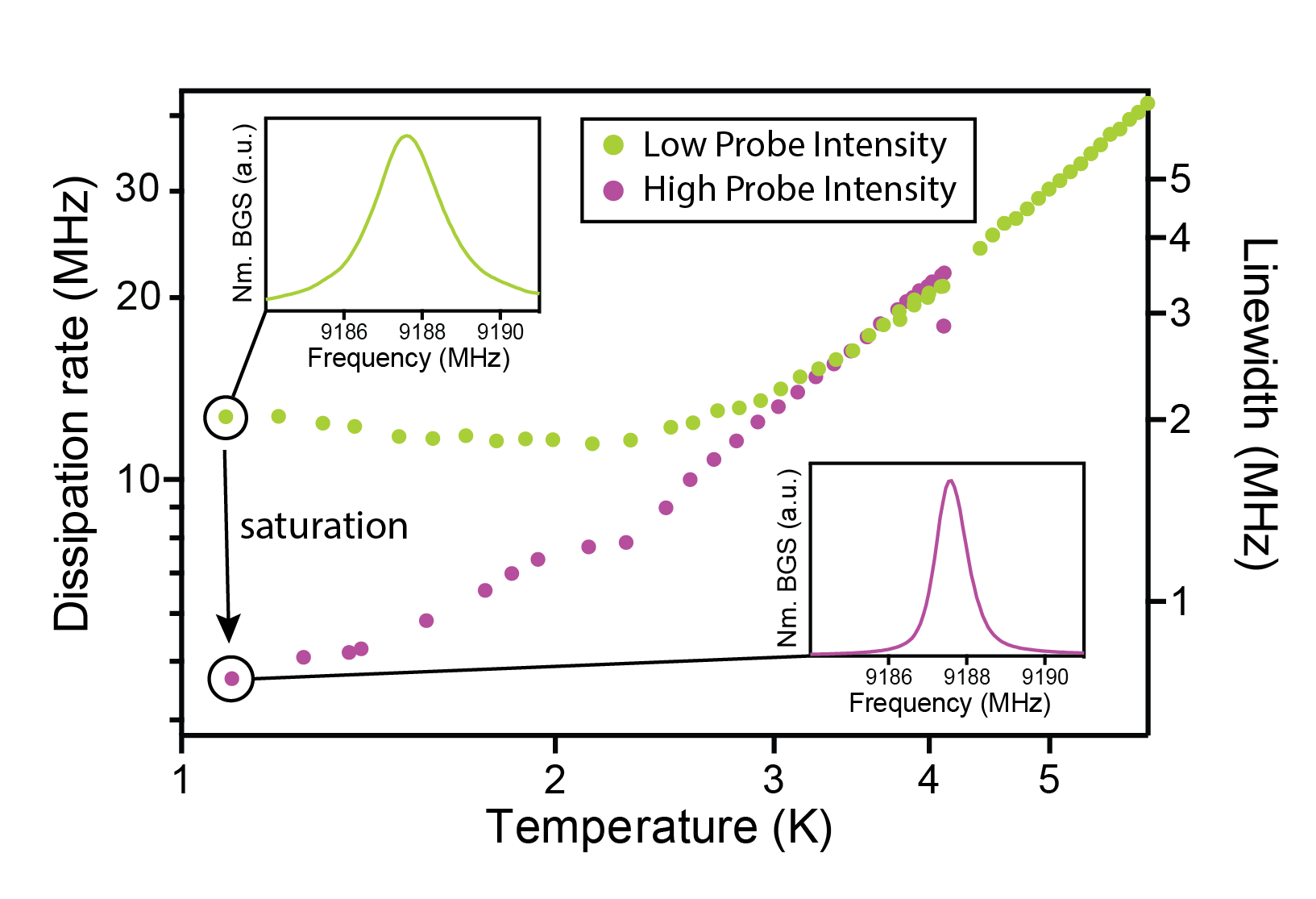}
\caption{Temperature dependence of acoustic dissipation in small core fiber. For low probe intensities (green points) the dissipation begins to rise with lower temperature, indicating the onset of resonant absorption by acoustic atoms. Resonant absorption can be saturated by increasing the intensity of the probe (purple points). Insets: Narrowing of the acoustic linewidth with increased probe intensity is evident in the normalized Brillouin gain spectra (Nm. BGS).}
\label{TLSevidence}
\end{center}
\end{figure}

In a previous study we measured the temperature and intensity dependence of acoustic dissipation in small core fiber \cite{Behunin2016}. In these experiments the system was studied by variation of the probe intensity, no drive was employed. These measurements showed that acoustic dissipation begins to rise with lower temperature near 2 K (green data points Fig.\ref{TLSevidence}). This behavior is indicative of resonant absorption of phonons by acoustic atoms. Acoustic atoms absorb sound in manner analogous to the way atoms absorb light. Thus, acoustic atoms in their ground state can absorb phonons with energies matching their gap $E$. Therefore this absorption mechanism will become important at low temperatures when the defects that interact with the 9 GHz phonons that we study relax to their ground states. These defects relax to their ground state near $0.5$ K, but a majority are in their ground state at the 1-2 K temperatures achieved of our apparatus.  

The two-level nature of this absorption mechanism can be verified by increasing the phonon intensity. High intensity acoustic fields drive acoustic atoms into transparency and eliminate this source of dissipation. This is clearly demonstrated by the purple data points of Fig. \ref{TLSevidence} which show that the dissipation rate is lowered as the probe intensity is increased. Two representative Brillouin gain spectra show how the energy transfer bandwidth narrows at high probe intensity.  

\subsection{TLS-phonon coupling}
In this section we show how our model of spectral hole burning can be extended to other structures. The extension of the model to other structures is accounted for in the model coupling $g_{q,j}$. For acoustic modes that are not plane waves the defect-phonon coupling  rate is generally given by 
\begin{equation}
\label{ }
g_{q,j} = 
\frac{\Delta_{0,j}}{E_j} \sqrt{\frac{1}{2\hbar \Omega_q}} \gamma : \underline{\xi}_q(x_j)
\end{equation}
where $\gamma : \underline{\xi}_q \equiv \gamma^{ab}\xi_{q,ab}$ is the tensor product between the orientation dependent deformation potential tensor and the elastic strain eigenfunction defined by $\xi_{q,ab} = (\partial_a u_{q,b} + \partial_b u_{q,a})/2$ where $u_{q,a}$ is the $a$th component of the elastic displacement eigenfunction defined by $\partial_i C^{ijkl} \partial_k u_{q,l} = - \rho \Omega_q^2 u^j_q$ and $\int d^3 x \ \rho u^*_q u_{q'} = \delta_{qq'}$ ($C^{ijkl}$ is the elastic tensor)
\cite{Behunin2016b}. In the plane wave limit $\gamma : \xi_q = \gamma q/\sqrt{\rho V}$ where $q$ is the phonon wavevector and $V$ is the system volume, giving the expression  
$|g_q|^2 = (\Delta_0/E)^2 (1/2\hbar)\gamma^2 \Omega_q/\rho V$.

\bibliography{/Users/rbehunin/sync/Bibtex/HoleBurning}

\begin{thebibliography}{53}%
\makeatletter
\providecommand \@ifxundefined [1]{%
 \@ifx{#1\undefined}
}%
\providecommand \@ifnum [1]{%
 \ifnum #1\expandafter \@firstoftwo
 \else \expandafter \@secondoftwo
 \fi
}%
\providecommand \@ifx [1]{%
 \ifx #1\expandafter \@firstoftwo
 \else \expandafter \@secondoftwo
 \fi
}%
\providecommand \natexlab [1]{#1}%
\providecommand \enquote  [1]{``#1''}%
\providecommand \bibnamefont  [1]{#1}%
\providecommand \bibfnamefont [1]{#1}%
\providecommand \citenamefont [1]{#1}%
\providecommand \href@noop [0]{\@secondoftwo}%
\providecommand \href [0]{\begingroup \@sanitize@url \@href}%
\providecommand \@href[1]{\@@startlink{#1}\@@href}%
\providecommand \@@href[1]{\endgroup#1\@@endlink}%
\providecommand \@sanitize@url [0]{\catcode `\\12\catcode `\$12\catcode
  `\&12\catcode `\#12\catcode `\^12\catcode `\_12\catcode `\%12\relax}%
\providecommand \@@startlink[1]{}%
\providecommand \@@endlink[0]{}%
\providecommand \url  [0]{\begingroup\@sanitize@url \@url }%
\providecommand \@url [1]{\endgroup\@href {#1}{\urlprefix }}%
\providecommand \urlprefix  [0]{URL }%
\providecommand \Eprint [0]{\href }%
\providecommand \doibase [0]{http://dx.doi.org/}%
\providecommand \selectlanguage [0]{\@gobble}%
\providecommand \bibinfo  [0]{\@secondoftwo}%
\providecommand \bibfield  [0]{\@secondoftwo}%
\providecommand \translation [1]{[#1]}%
\providecommand \BibitemOpen [0]{}%
\providecommand \bibitemStop [0]{}%
\providecommand \bibitemNoStop [0]{.\EOS\space}%
\providecommand \EOS [0]{\spacefactor3000\relax}%
\providecommand \BibitemShut  [1]{\csname bibitem#1\endcsname}%
\let\auto@bib@innerbib\@empty
\bibitem [{\citenamefont {Kippenberg}\ and\ \citenamefont
  {Vahala}(2008)}]{Kippenberg2008}%
  \BibitemOpen
  \bibfield  {author} {\bibinfo {author} {\bibfnamefont {T.~J.}\ \bibnamefont
  {Kippenberg}}\ and\ \bibinfo {author} {\bibfnamefont {K.~J.}\ \bibnamefont
  {Vahala}},\ }\href {\doibase 10.1126/science.1156032} {\bibfield  {journal}
  {\bibinfo  {journal} {Science}\ }\textbf {\bibinfo {volume} {321}},\ \bibinfo
  {pages} {1172} (\bibinfo {year} {2008})}\BibitemShut {NoStop}%
\bibitem [{\citenamefont {Regal}\ \emph {et~al.}(2008)\citenamefont {Regal},
  \citenamefont {Teufel},\ and\ \citenamefont {Lehnert}}]{Regal2008}%
  \BibitemOpen
  \bibfield  {author} {\bibinfo {author} {\bibfnamefont {C.~A.}\ \bibnamefont
  {Regal}}, \bibinfo {author} {\bibfnamefont {J.~D.}\ \bibnamefont {Teufel}}, \
  and\ \bibinfo {author} {\bibfnamefont {K.~W.}\ \bibnamefont {Lehnert}},\
  }\href@noop {} {\bibfield  {journal} {\bibinfo  {journal} {Nature Physics}\
  ,\ \bibinfo {pages} {7}} (\bibinfo {year} {2008})}\BibitemShut {NoStop}%
\bibitem [{\citenamefont {Bochmann}\ \emph {et~al.}(2013)\citenamefont
  {Bochmann}, \citenamefont {Vainsencher}, \citenamefont {Awschalom},\ and\
  \citenamefont {Cleland}}]{Bochmann2013}%
  \BibitemOpen
  \bibfield  {author} {\bibinfo {author} {\bibfnamefont {J.}~\bibnamefont
  {Bochmann}}, \bibinfo {author} {\bibfnamefont {A.}~\bibnamefont
  {Vainsencher}}, \bibinfo {author} {\bibfnamefont {D.~D.}\ \bibnamefont
  {Awschalom}}, \ and\ \bibinfo {author} {\bibfnamefont {A.~N.}\ \bibnamefont
  {Cleland}},\ }\href {\doibase 10.1038/nphys2748} {\bibfield  {journal}
  {\bibinfo  {journal} {Nature Physics}\ }\textbf {\bibinfo {volume} {9}},\
  \bibinfo {pages} {712} (\bibinfo {year} {2013})}\BibitemShut {NoStop}%
\bibitem [{\citenamefont {Mahboob}\ \emph {et~al.}(2013)\citenamefont
  {Mahboob}, \citenamefont {Nishiguchi}, \citenamefont {Fujiwara},\ and\
  \citenamefont {Yamaguchi}}]{Mahboob2013}%
  \BibitemOpen
  \bibfield  {author} {\bibinfo {author} {\bibfnamefont {I.}~\bibnamefont
  {Mahboob}}, \bibinfo {author} {\bibfnamefont {K.}~\bibnamefont {Nishiguchi}},
  \bibinfo {author} {\bibfnamefont {A.}~\bibnamefont {Fujiwara}}, \ and\
  \bibinfo {author} {\bibfnamefont {H.}~\bibnamefont {Yamaguchi}},\ }\href
  {\doibase 10.1103/PhysRevLett.110.127202} {\bibfield  {journal} {\bibinfo
  {journal} {Physical Review Letters}\ }\textbf {\bibinfo {volume} {110}},\
  \bibinfo {pages} {127202} (\bibinfo {year} {2013})}\BibitemShut {NoStop}%
\bibitem [{\citenamefont {Schliesser}\ \emph {et~al.}(2009)\citenamefont
  {Schliesser}, \citenamefont {Arcizet}, \citenamefont {Rivi{\`{e}}re},
  \citenamefont {Anetsberger},\ and\ \citenamefont
  {Kippenberg}}]{Schliesser2009}%
  \BibitemOpen
  \bibfield  {author} {\bibinfo {author} {\bibfnamefont {A.}~\bibnamefont
  {Schliesser}}, \bibinfo {author} {\bibfnamefont {O.}~\bibnamefont {Arcizet}},
  \bibinfo {author} {\bibfnamefont {R.}~\bibnamefont {Rivi{\`{e}}re}}, \bibinfo
  {author} {\bibfnamefont {G.}~\bibnamefont {Anetsberger}}, \ and\ \bibinfo
  {author} {\bibfnamefont {T.~J.}\ \bibnamefont {Kippenberg}},\ }\href
  {\doibase 10.1038/nphys1304} {\bibfield  {journal} {\bibinfo  {journal}
  {Nature Physics}\ }\textbf {\bibinfo {volume} {5}},\ \bibinfo {pages} {509}
  (\bibinfo {year} {2009})}\BibitemShut {NoStop}%
\bibitem [{\citenamefont {Lee}\ \emph {et~al.}(2012)\citenamefont {Lee},
  \citenamefont {Chen}, \citenamefont {Li}, \citenamefont {Yang}, \citenamefont
  {Jeon}, \citenamefont {Painter},\ and\ \citenamefont {Vahala}}]{Lee2012}%
  \BibitemOpen
  \bibfield  {author} {\bibinfo {author} {\bibfnamefont {H.}~\bibnamefont
  {Lee}}, \bibinfo {author} {\bibfnamefont {T.}~\bibnamefont {Chen}}, \bibinfo
  {author} {\bibfnamefont {J.}~\bibnamefont {Li}}, \bibinfo {author}
  {\bibfnamefont {K.~Y.}\ \bibnamefont {Yang}}, \bibinfo {author}
  {\bibfnamefont {S.}~\bibnamefont {Jeon}}, \bibinfo {author} {\bibfnamefont
  {O.}~\bibnamefont {Painter}}, \ and\ \bibinfo {author} {\bibfnamefont
  {K.~J.}\ \bibnamefont {Vahala}},\ }\href@noop {} {\bibfield  {journal}
  {\bibinfo  {journal} {Nature Photonics}\ }\textbf {\bibinfo {volume} {6}},\
  \bibinfo {pages} {369} (\bibinfo {year} {2012})}\BibitemShut {NoStop}%
\bibitem [{\citenamefont {Goryachev}\ \emph {et~al.}(2013)\citenamefont
  {Goryachev}, \citenamefont {Creedon}, \citenamefont {Galliou},\ and\
  \citenamefont {Tobar}}]{Goryachev2013d}%
  \BibitemOpen
  \bibfield  {author} {\bibinfo {author} {\bibfnamefont {M.}~\bibnamefont
  {Goryachev}}, \bibinfo {author} {\bibfnamefont {D.~L.}\ \bibnamefont
  {Creedon}}, \bibinfo {author} {\bibfnamefont {S.}~\bibnamefont {Galliou}}, \
  and\ \bibinfo {author} {\bibfnamefont {M.~E.}\ \bibnamefont {Tobar}},\ }\href
  {\doibase 10.1103/PhysRevLett.111.085502} {\bibfield  {journal} {\bibinfo
  {journal} {Physical Review Letters}\ }\textbf {\bibinfo {volume} {111}},\
  \bibinfo {pages} {085502} (\bibinfo {year} {2013})}\BibitemShut {NoStop}%
\bibitem [{\citenamefont {Meenehan}\ \emph {et~al.}(2015)\citenamefont
  {Meenehan}, \citenamefont {Cohen}, \citenamefont {MacCabe}, \citenamefont
  {Marsili}, \citenamefont {Shaw},\ and\ \citenamefont
  {Painter}}]{Meenehan2015a}%
  \BibitemOpen
  \bibfield  {author} {\bibinfo {author} {\bibfnamefont {S.~M.}\ \bibnamefont
  {Meenehan}}, \bibinfo {author} {\bibfnamefont {J.~D.}\ \bibnamefont {Cohen}},
  \bibinfo {author} {\bibfnamefont {G.~S.}\ \bibnamefont {MacCabe}}, \bibinfo
  {author} {\bibfnamefont {F.}~\bibnamefont {Marsili}}, \bibinfo {author}
  {\bibfnamefont {M.~D.}\ \bibnamefont {Shaw}}, \ and\ \bibinfo {author}
  {\bibfnamefont {O.}~\bibnamefont {Painter}},\ }\href {\doibase
  10.1103/PhysRevX.5.041002} {\bibfield  {journal} {\bibinfo  {journal}
  {Physical Review X}\ }\textbf {\bibinfo {volume} {5}},\ \bibinfo {pages} {1}
  (\bibinfo {year} {2015})}\BibitemShut {NoStop}%
\bibitem [{\citenamefont {Phillips}(1987)}]{Phillips1987}%
  \BibitemOpen
  \bibfield  {author} {\bibinfo {author} {\bibfnamefont {W.~A.}\ \bibnamefont
  {Phillips}},\ }\href {\doibase 10.1088/0034-4885/50/12/003} {\bibfield
  {journal} {\bibinfo  {journal} {Reports on Progress in Physics}\ }\textbf
  {\bibinfo {volume} {50}},\ \bibinfo {pages} {1657} (\bibinfo {year}
  {1987})}\BibitemShut {NoStop}%
\bibitem [{\citenamefont {Shin}\ \emph {et~al.}(2013)\citenamefont {Shin},
  \citenamefont {Qiu}, \citenamefont {Jarecki}, \citenamefont {Cox},
  \citenamefont {Olsson}, \citenamefont {Starbuck}, \citenamefont {Wang},\ and\
  \citenamefont {Rakich}}]{Shin2013}%
  \BibitemOpen
  \bibfield  {author} {\bibinfo {author} {\bibfnamefont {H.}~\bibnamefont
  {Shin}}, \bibinfo {author} {\bibfnamefont {W.}~\bibnamefont {Qiu}}, \bibinfo
  {author} {\bibfnamefont {R.}~\bibnamefont {Jarecki}}, \bibinfo {author}
  {\bibfnamefont {J.~A.}\ \bibnamefont {Cox}}, \bibinfo {author} {\bibfnamefont
  {R.~H.}\ \bibnamefont {Olsson}}, \bibinfo {author} {\bibfnamefont
  {A.}~\bibnamefont {Starbuck}}, \bibinfo {author} {\bibfnamefont
  {Z.}~\bibnamefont {Wang}}, \ and\ \bibinfo {author} {\bibfnamefont {P.~T.}\
  \bibnamefont {Rakich}},\ }\href {\doibase 10.1038/ncomms2943} {\bibfield
  {journal} {\bibinfo  {journal} {Nature communications}\ }\textbf {\bibinfo
  {volume} {4}},\ \bibinfo {pages} {1944} (\bibinfo {year} {2013})}\BibitemShut
  {NoStop}%
\bibitem [{\citenamefont {Shin}\ \emph {et~al.}(2015)\citenamefont {Shin},
  \citenamefont {Cox}, \citenamefont {Jarecki}, \citenamefont {Starbuck},
  \citenamefont {Wang},\ and\ \citenamefont {Rakich}}]{Shin2015}%
  \BibitemOpen
  \bibfield  {author} {\bibinfo {author} {\bibfnamefont {H.}~\bibnamefont
  {Shin}}, \bibinfo {author} {\bibfnamefont {J.~A.}\ \bibnamefont {Cox}},
  \bibinfo {author} {\bibfnamefont {R.}~\bibnamefont {Jarecki}}, \bibinfo
  {author} {\bibfnamefont {A.}~\bibnamefont {Starbuck}}, \bibinfo {author}
  {\bibfnamefont {Z.}~\bibnamefont {Wang}}, \ and\ \bibinfo {author}
  {\bibfnamefont {P.~T.}\ \bibnamefont {Rakich}},\ }\href {\doibase
  10.1038/ncomms7427} {\bibfield  {journal} {\bibinfo  {journal} {Nature
  communications}\ }\textbf {\bibinfo {volume} {6}},\ \bibinfo {pages} {6427}
  (\bibinfo {year} {2015})}\BibitemShut {NoStop}%
\bibitem [{\citenamefont {Chan}\ \emph {et~al.}(2011)\citenamefont {Chan},
  \citenamefont {Alegre}, \citenamefont {Safavi-Naeini}, \citenamefont {Hill},
  \citenamefont {Krause}, \citenamefont {Groeblacher}, \citenamefont
  {Aspelmeyer},\ and\ \citenamefont {Painter}}]{Chan2011}%
  \BibitemOpen
  \bibfield  {author} {\bibinfo {author} {\bibfnamefont {J.}~\bibnamefont
  {Chan}}, \bibinfo {author} {\bibfnamefont {T.~P.~M.}\ \bibnamefont {Alegre}},
  \bibinfo {author} {\bibfnamefont {A.~H.}\ \bibnamefont {Safavi-Naeini}},
  \bibinfo {author} {\bibfnamefont {J.~T.}\ \bibnamefont {Hill}}, \bibinfo
  {author} {\bibfnamefont {A.}~\bibnamefont {Krause}}, \bibinfo {author}
  {\bibfnamefont {S.}~\bibnamefont {Groeblacher}}, \bibinfo {author}
  {\bibfnamefont {M.}~\bibnamefont {Aspelmeyer}}, \ and\ \bibinfo {author}
  {\bibfnamefont {O.}~\bibnamefont {Painter}},\ }\href {\doibase
  10.1038/nature10461} {\bibfield  {journal} {\bibinfo  {journal} {Nature}\
  }\textbf {\bibinfo {volume} {478}},\ \bibinfo {pages} {18} (\bibinfo {year}
  {2011})}\BibitemShut {NoStop}%
\bibitem [{\citenamefont {Fiore}\ \emph {et~al.}(2011)\citenamefont {Fiore},
  \citenamefont {Yang}, \citenamefont {Kuzyk}, \citenamefont {Barbour},
  \citenamefont {Tian},\ and\ \citenamefont {Wang}}]{Fiore2011}%
  \BibitemOpen
  \bibfield  {author} {\bibinfo {author} {\bibfnamefont {V.}~\bibnamefont
  {Fiore}}, \bibinfo {author} {\bibfnamefont {Y.}~\bibnamefont {Yang}},
  \bibinfo {author} {\bibfnamefont {M.~C.}\ \bibnamefont {Kuzyk}}, \bibinfo
  {author} {\bibfnamefont {R.}~\bibnamefont {Barbour}}, \bibinfo {author}
  {\bibfnamefont {L.}~\bibnamefont {Tian}}, \ and\ \bibinfo {author}
  {\bibfnamefont {H.}~\bibnamefont {Wang}},\ }\href {\doibase
  10.1103/PhysRevLett.107.133601} {\bibfield  {journal} {\bibinfo  {journal}
  {Physical Review Letters}\ }\textbf {\bibinfo {volume} {107}},\ \bibinfo
  {pages} {1} (\bibinfo {year} {2011})}\BibitemShut {NoStop}%
\bibitem [{\citenamefont {Weis}\ \emph {et~al.}(2010)\citenamefont {Weis},
  \citenamefont {Riviere}, \citenamefont {Deleglise}, \citenamefont {Gavartin},
  \citenamefont {Arcizet}, \citenamefont {Schliesser},\ and\ \citenamefont
  {Kippenberg}}]{Weis2010}%
  \BibitemOpen
  \bibfield  {author} {\bibinfo {author} {\bibfnamefont {S.}~\bibnamefont
  {Weis}}, \bibinfo {author} {\bibfnamefont {R.}~\bibnamefont {Riviere}},
  \bibinfo {author} {\bibfnamefont {S.}~\bibnamefont {Deleglise}}, \bibinfo
  {author} {\bibfnamefont {E.}~\bibnamefont {Gavartin}}, \bibinfo {author}
  {\bibfnamefont {O.}~\bibnamefont {Arcizet}}, \bibinfo {author} {\bibfnamefont
  {A.}~\bibnamefont {Schliesser}}, \ and\ \bibinfo {author} {\bibfnamefont
  {T.~J.}\ \bibnamefont {Kippenberg}},\ }\href {\doibase
  10.1126/science.1195596} {\bibfield  {journal} {\bibinfo  {journal}
  {Science}\ }\textbf {\bibinfo {volume} {330}},\ \bibinfo {pages} {1520}
  (\bibinfo {year} {2010})}\BibitemShut {NoStop}%
\bibitem [{\citenamefont {Cheung}\ \emph {et~al.}()\citenamefont {Cheung},
  \citenamefont {Patil}, \citenamefont {Chang}, \citenamefont {Chakram},\ and\
  \citenamefont {Vengalattore}}]{Cheung2016}%
  \BibitemOpen
  \bibfield  {author} {\bibinfo {author} {\bibfnamefont {H.~F.~H.}\
  \bibnamefont {Cheung}}, \bibinfo {author} {\bibfnamefont {Y.~S.}\
  \bibnamefont {Patil}}, \bibinfo {author} {\bibfnamefont {L.}~\bibnamefont
  {Chang}}, \bibinfo {author} {\bibfnamefont {S.}~\bibnamefont {Chakram}}, \
  and\ \bibinfo {author} {\bibfnamefont {M.}~\bibnamefont {Vengalattore}},\
  }\href {http://arxiv.org/abs/1601.02324} {\ }\Eprint
  {http://arxiv.org/abs/1601.02324} {arXiv:1601.02324} \BibitemShut {NoStop}%
\bibitem [{\citenamefont {Kang}\ \emph {et~al.}(2009)\citenamefont {Kang},
  \citenamefont {Nazarkin}, \citenamefont {Brenn},\ and\ \citenamefont
  {Russell}}]{Kang2009}%
  \BibitemOpen
  \bibfield  {author} {\bibinfo {author} {\bibfnamefont {M.~S.}\ \bibnamefont
  {Kang}}, \bibinfo {author} {\bibfnamefont {A.}~\bibnamefont {Nazarkin}},
  \bibinfo {author} {\bibfnamefont {A.}~\bibnamefont {Brenn}}, \ and\ \bibinfo
  {author} {\bibfnamefont {P.~S.~J.}\ \bibnamefont {Russell}},\ }\href
  {\doibase 10.1038/nphys1217} {\bibfield  {journal} {\bibinfo  {journal}
  {Nature Physics}\ }\textbf {\bibinfo {volume} {5}},\ \bibinfo {pages} {276}
  (\bibinfo {year} {2009})}\BibitemShut {NoStop}%
\bibitem [{\citenamefont {Schliesser}\ \emph {et~al.}(2007)\citenamefont
  {Schliesser}, \citenamefont {Arcizet}, \citenamefont {Wilken}, \citenamefont
  {Holzwarth}, \citenamefont {Kippenberg}, \citenamefont {Del'Haye},\ and\
  \citenamefont {Haye}}]{DelHaye2007}%
  \BibitemOpen
  \bibfield  {author} {\bibinfo {author} {\bibfnamefont {A.}~\bibnamefont
  {Schliesser}}, \bibinfo {author} {\bibfnamefont {O.}~\bibnamefont {Arcizet}},
  \bibinfo {author} {\bibfnamefont {T.}~\bibnamefont {Wilken}}, \bibinfo
  {author} {\bibfnamefont {R.}~\bibnamefont {Holzwarth}}, \bibinfo {author}
  {\bibfnamefont {T.~J.}\ \bibnamefont {Kippenberg}}, \bibinfo {author}
  {\bibfnamefont {P.}~\bibnamefont {Del'Haye}}, \ and\ \bibinfo {author}
  {\bibfnamefont {P.~D.}\ \bibnamefont {Haye}},\ }\href {\doibase
  10.1038/nature06401} {\bibfield  {journal} {\bibinfo  {journal} {Nature}\
  }\textbf {\bibinfo {volume} {450}},\ \bibinfo {pages} {1214} (\bibinfo {year}
  {2007})}\BibitemShut {NoStop}%
\bibitem [{\citenamefont {Kippenberg}\ \emph {et~al.}(2011)\citenamefont
  {Kippenberg}, \citenamefont {Holzwarth},\ and\ \citenamefont
  {Diddams}}]{Kippenberg2011}%
  \BibitemOpen
  \bibfield  {author} {\bibinfo {author} {\bibfnamefont {T.~J.}\ \bibnamefont
  {Kippenberg}}, \bibinfo {author} {\bibfnamefont {R.}~\bibnamefont
  {Holzwarth}}, \ and\ \bibinfo {author} {\bibfnamefont {S.~A.}\ \bibnamefont
  {Diddams}},\ }\href {\doibase 10.1126/science.1193968} {\bibfield  {journal}
  {\bibinfo  {journal} {Science}\ }\textbf {\bibinfo {volume} {332}},\ \bibinfo
  {pages} {555} (\bibinfo {year} {2011})}\BibitemShut {NoStop}%
\bibitem [{\citenamefont {Portis}(1953)}]{Portis1953}%
  \BibitemOpen
  \bibfield  {author} {\bibinfo {author} {\bibfnamefont {A.~M.}\ \bibnamefont
  {Portis}},\ }\href@noop {} {\bibfield  {journal} {\bibinfo  {journal}
  {Physical Review}\ }\textbf {\bibinfo {volume} {91}},\ \bibinfo {pages}
  {1071} (\bibinfo {year} {1953})}\BibitemShut {NoStop}%
\bibitem [{\citenamefont {Szabo}(1975)}]{Szabo1975}%
  \BibitemOpen
  \bibfield  {author} {\bibinfo {author} {\bibfnamefont {A.}~\bibnamefont
  {Szabo}},\ }\href@noop {} {\bibfield  {journal} {\bibinfo  {journal}
  {Physical Review B}\ }\textbf {\bibinfo {volume} {11}},\ \bibinfo {pages}
  {4512} (\bibinfo {year} {1975})}\BibitemShut {NoStop}%
\bibitem [{\citenamefont {Jessop}\ \emph {et~al.}(1980)\citenamefont {Jessop},
  \citenamefont {Muramoto},\ and\ \citenamefont {Szabo}}]{Jessop1980}%
  \BibitemOpen
  \bibfield  {author} {\bibinfo {author} {\bibfnamefont {P.}~\bibnamefont
  {Jessop}}, \bibinfo {author} {\bibfnamefont {T.}~\bibnamefont {Muramoto}}, \
  and\ \bibinfo {author} {\bibfnamefont {A.}~\bibnamefont {Szabo}},\
  }\href@noop {} {\bibfield  {journal} {\bibinfo  {journal} {Physical Review
  B}\ }\textbf {\bibinfo {volume} {21}},\ \bibinfo {pages} {926} (\bibinfo
  {year} {1980})}\BibitemShut {NoStop}%
\bibitem [{\citenamefont {Basche}\ and\ \citenamefont
  {Moerner}(1992)}]{Basche1992}%
  \BibitemOpen
  \bibfield  {author} {\bibinfo {author} {\bibfnamefont {T.}~\bibnamefont
  {Basche}}\ and\ \bibinfo {author} {\bibfnamefont {W.~E.}\ \bibnamefont
  {Moerner}},\ }\href {\doibase 10.1038/355335a0} {\bibfield  {journal}
  {\bibinfo  {journal} {Nature}\ }\textbf {\bibinfo {volume} {355}},\ \bibinfo
  {pages} {335} (\bibinfo {year} {1992})}\BibitemShut {NoStop}%
\bibitem [{\citenamefont {Golding}\ \emph {et~al.}(1973)\citenamefont
  {Golding}, \citenamefont {Graebner}, \citenamefont {Halperin},\ and\
  \citenamefont {Schutz}}]{Golding1973}%
  \BibitemOpen
  \bibfield  {author} {\bibinfo {author} {\bibfnamefont {B.}~\bibnamefont
  {Golding}}, \bibinfo {author} {\bibfnamefont {J.~E.}\ \bibnamefont
  {Graebner}}, \bibinfo {author} {\bibfnamefont {B.~I.}\ \bibnamefont
  {Halperin}}, \ and\ \bibinfo {author} {\bibfnamefont {R.~J.}\ \bibnamefont
  {Schutz}},\ }\href@noop {} {\bibfield  {journal} {\bibinfo  {journal}
  {Physical Review Letters}\ }\textbf {\bibinfo {volume} {30}},\ \bibinfo
  {pages} {223} (\bibinfo {year} {1973})}\BibitemShut {NoStop}%
\bibitem [{\citenamefont {Golding}\ \emph {et~al.}(1976)\citenamefont
  {Golding}, \citenamefont {Graebner},\ and\ \citenamefont
  {Schutz}}]{Golding1976b}%
  \BibitemOpen
  \bibfield  {author} {\bibinfo {author} {\bibfnamefont {B.}~\bibnamefont
  {Golding}}, \bibinfo {author} {\bibfnamefont {J.~E.}\ \bibnamefont
  {Graebner}}, \ and\ \bibinfo {author} {\bibfnamefont {R.~J.}\ \bibnamefont
  {Schutz}},\ }\href@noop {} {\bibfield  {journal} {\bibinfo  {journal}
  {Physical Review B}\ }\textbf {\bibinfo {volume} {14}},\ \bibinfo {pages} {0}
  (\bibinfo {year} {1976})}\BibitemShut {NoStop}%
\bibitem [{\citenamefont {Arnold}\ \emph {et~al.}(1974)\citenamefont {Arnold},
  \citenamefont {Hunklinger}, \citenamefont {Stein},\ and\ \citenamefont
  {Dransfeld}}]{Arnold1974}%
  \BibitemOpen
  \bibfield  {author} {\bibinfo {author} {\bibfnamefont {W.}~\bibnamefont
  {Arnold}}, \bibinfo {author} {\bibfnamefont {S.}~\bibnamefont {Hunklinger}},
  \bibinfo {author} {\bibfnamefont {S.}~\bibnamefont {Stein}}, \ and\ \bibinfo
  {author} {\bibfnamefont {K.}~\bibnamefont {Dransfeld}},\ }\href@noop {}
  {\bibfield  {journal} {\bibinfo  {journal} {Jounal of Non-Crystalline
  Solids}\ }\textbf {\bibinfo {volume} {14}},\ \bibinfo {pages} {192} (\bibinfo
  {year} {1974})}\BibitemShut {NoStop}%
\bibitem [{\citenamefont {Arnold}\ and\ \citenamefont
  {Hunklinger}(1975)}]{Arnold1975}%
  \BibitemOpen
  \bibfield  {author} {\bibinfo {author} {\bibfnamefont {W.}~\bibnamefont
  {Arnold}}\ and\ \bibinfo {author} {\bibfnamefont {S.}~\bibnamefont
  {Hunklinger}},\ }\href {\doibase 10.1017/CBO9781107415324.004} {\bibfield
  {journal} {\bibinfo  {journal} {Solid State Communications}\ }\textbf
  {\bibinfo {volume} {17}},\ \bibinfo {pages} {883} (\bibinfo {year}
  {1975})}\BibitemShut {NoStop}%
\bibitem [{\citenamefont {Arnold}\ \emph {et~al.}(1978)\citenamefont {Arnold},
  \citenamefont {Martinon},\ and\ \citenamefont {Hunklinger}}]{Arnold1978}%
  \BibitemOpen
  \bibfield  {author} {\bibinfo {author} {\bibfnamefont {W.}~\bibnamefont
  {Arnold}}, \bibinfo {author} {\bibfnamefont {C.}~\bibnamefont {Martinon}}, \
  and\ \bibinfo {author} {\bibfnamefont {S.}~\bibnamefont {Hunklinger}},\
  }\href@noop {} {\bibfield  {journal} {\bibinfo  {journal} {J. Phys.
  Colloques}\ }\textbf {\bibinfo {volume} {39}},\ \bibinfo {pages} {C6961}
  (\bibinfo {year} {1978})}\BibitemShut {NoStop}%
\bibitem [{\citenamefont {Black}\ and\ \citenamefont
  {Halperin}(1977)}]{Black1977}%
  \BibitemOpen
  \bibfield  {author} {\bibinfo {author} {\bibfnamefont {J.~L.}\ \bibnamefont
  {Black}}\ and\ \bibinfo {author} {\bibfnamefont {B.~I.}\ \bibnamefont
  {Halperin}},\ }\href {\doibase 10.1016/0038-1098(77)91219-4} {\bibfield
  {journal} {\bibinfo  {journal} {Physical Review B}\ }\textbf {\bibinfo
  {volume} {24}},\ \bibinfo {pages} {813} (\bibinfo {year} {1977})}\BibitemShut
  {NoStop}%
\bibitem [{\citenamefont {Behunin}\ \emph {et~al.}(2016)\citenamefont
  {Behunin}, \citenamefont {Intravaia},\ and\ \citenamefont
  {Rakich}}]{Behunin2016b}%
  \BibitemOpen
  \bibfield  {author} {\bibinfo {author} {\bibfnamefont {R.~O.}\ \bibnamefont
  {Behunin}}, \bibinfo {author} {\bibfnamefont {F.}~\bibnamefont {Intravaia}},
  \ and\ \bibinfo {author} {\bibfnamefont {P.~T.}\ \bibnamefont {Rakich}},\
  }\href {http://arxiv.org/abs/1601.06837} {\bibfield  {journal} {\bibinfo
  {journal} {Physical Review B}\ }\textbf {\bibinfo {volume} {XX}},\ \bibinfo
  {pages} {XXXXX} (\bibinfo {year} {2016})},\ \Eprint
  {http://arxiv.org/abs/1601.06837} {arXiv:1601.06837} \BibitemShut {NoStop}%
\bibitem [{\citenamefont {Siegman}(1986)}]{Siegman1986}%
  \BibitemOpen
  \bibfield  {author} {\bibinfo {author} {\bibfnamefont {A.~E.}\ \bibnamefont
  {Siegman}},\ }\href@noop {} {\emph {\bibinfo {title} {{Lasers}}}}\ (\bibinfo
  {publisher} {University Science Books},\ \bibinfo {address} {Mill Valley,
  CA},\ \bibinfo {year} {1986})\BibitemShut {NoStop}%
\bibitem [{\citenamefont {Skinner}\ and\ \citenamefont
  {Moerner}(1996)}]{Skinner1996}%
  \BibitemOpen
  \bibfield  {author} {\bibinfo {author} {\bibfnamefont {J.~L.}\ \bibnamefont
  {Skinner}}\ and\ \bibinfo {author} {\bibfnamefont {W.~E.}\ \bibnamefont
  {Moerner}},\ }\href {\doibase 10.1021/jp9601328} {\bibfield  {journal}
  {\bibinfo  {journal} {Journal of Physical Chemistry}\ }\textbf {\bibinfo
  {volume} {100}},\ \bibinfo {pages} {13251} (\bibinfo {year}
  {1996})}\BibitemShut {NoStop}%
\bibitem [{\citenamefont {Jankowiak}\ and\ \citenamefont
  {Small}(1987)}]{Jankowiak1987}%
  \BibitemOpen
  \bibfield  {author} {\bibinfo {author} {\bibfnamefont {R.}~\bibnamefont
  {Jankowiak}}\ and\ \bibinfo {author} {\bibfnamefont {G.~J.}\ \bibnamefont
  {Small}},\ }\href@noop {} {\bibfield  {journal} {\bibinfo  {journal}
  {Science}\ }\textbf {\bibinfo {volume} {237}},\ \bibinfo {pages} {618}
  (\bibinfo {year} {1987})}\BibitemShut {NoStop}%
\bibitem [{\citenamefont {{Bjorklund, G.C., Haarer, D., Hayes, J.M., Jankowiak,
  R., Lenth, W., Macfarlane, R.M., Rebane, K.K., Rebane, L.A., Shelby, R.M.,
  Sievers, A.J. and Small}}(1988)}]{Bjorklund2012}%
  \BibitemOpen
  \bibfield  {author} {\bibinfo {author} {\bibfnamefont {G.}~\bibnamefont
  {{Bjorklund, G.C., Haarer, D., Hayes, J.M., Jankowiak, R., Lenth, W.,
  Macfarlane, R.M., Rebane, K.K., Rebane, L.A., Shelby, R.M., Sievers, A.J. and
  Small}}},\ }\href@noop {} {\emph {\bibinfo {title} {{Persistent Spectral
  Hole-Burning: Science and Applications}}}},\ edited by\ \bibinfo {editor}
  {\bibfnamefont {W.~E.}\ \bibnamefont {Moerner}}\ (\bibinfo  {publisher}
  {Springer},\ \bibinfo {address} {Berlin},\ \bibinfo {year}
  {1988})\BibitemShut {NoStop}%
\bibitem [{\citenamefont {Burnett}\ \emph {et~al.}(2014)\citenamefont
  {Burnett}, \citenamefont {Faoro}, \citenamefont {Wisby}, \citenamefont
  {Gurtovoi}, \citenamefont {Chernykh}, \citenamefont {Mikhailov},
  \citenamefont {Tulin}, \citenamefont {Shaikhaidarov}, \citenamefont
  {Antonov}, \citenamefont {Meeson}, \citenamefont {Tzalenchuk},\ and\
  \citenamefont {Lindstr{\"{o}}m}}]{Burnett2014a}%
  \BibitemOpen
  \bibfield  {author} {\bibinfo {author} {\bibfnamefont {J.}~\bibnamefont
  {Burnett}}, \bibinfo {author} {\bibfnamefont {L.}~\bibnamefont {Faoro}},
  \bibinfo {author} {\bibfnamefont {I.}~\bibnamefont {Wisby}}, \bibinfo
  {author} {\bibfnamefont {V.~L.}\ \bibnamefont {Gurtovoi}}, \bibinfo {author}
  {\bibfnamefont {A.~V.}\ \bibnamefont {Chernykh}}, \bibinfo {author}
  {\bibfnamefont {G.~M.}\ \bibnamefont {Mikhailov}}, \bibinfo {author}
  {\bibfnamefont {V.~A.}\ \bibnamefont {Tulin}}, \bibinfo {author}
  {\bibfnamefont {R.}~\bibnamefont {Shaikhaidarov}}, \bibinfo {author}
  {\bibfnamefont {V.}~\bibnamefont {Antonov}}, \bibinfo {author} {\bibfnamefont
  {P.~J.}\ \bibnamefont {Meeson}}, \bibinfo {author} {\bibfnamefont {a.~Y.}\
  \bibnamefont {Tzalenchuk}}, \ and\ \bibinfo {author} {\bibfnamefont
  {T.}~\bibnamefont {Lindstr{\"{o}}m}},\ }\href {\doibase 10.1038/ncomms5119}
  {\bibfield  {journal} {\bibinfo  {journal} {Nature communications}\ }\textbf
  {\bibinfo {volume} {5}},\ \bibinfo {pages} {4119} (\bibinfo {year}
  {2014})}\BibitemShut {NoStop}%
\bibitem [{\citenamefont {Boyd}(2003)}]{Boyd2003}%
  \BibitemOpen
  \bibfield  {author} {\bibinfo {author} {\bibfnamefont {R.~W.}\ \bibnamefont
  {Boyd}},\ }\href@noop {} {\emph {\bibinfo {title} {{Nonlinear Optics}}}},\
  \bibinfo {edition} {second edi}\ ed.\ (\bibinfo  {publisher} {Academic
  press},\ \bibinfo {year} {2003})\BibitemShut {NoStop}%
\bibitem [{\citenamefont {Behunin}\ \emph {et~al.}()\citenamefont {Behunin},
  \citenamefont {Kharel}, \citenamefont {Renninger}, \citenamefont {Shin},
  \citenamefont {Carter}, \citenamefont {Kittlaus},\ and\ \citenamefont
  {Rakich}}]{Behunin2016}%
  \BibitemOpen
  \bibfield  {author} {\bibinfo {author} {\bibfnamefont {R.~O.}\ \bibnamefont
  {Behunin}}, \bibinfo {author} {\bibfnamefont {P.}~\bibnamefont {Kharel}},
  \bibinfo {author} {\bibfnamefont {W.~H.}\ \bibnamefont {Renninger}}, \bibinfo
  {author} {\bibfnamefont {H.}~\bibnamefont {Shin}}, \bibinfo {author}
  {\bibfnamefont {F.}~\bibnamefont {Carter}}, \bibinfo {author} {\bibfnamefont
  {E.}~\bibnamefont {Kittlaus}}, \ and\ \bibinfo {author} {\bibfnamefont
  {P.~T.}\ \bibnamefont {Rakich}},\ }\href {http://arxiv.org/abs/1501.04248} {\
  }\Eprint {http://arxiv.org/abs/1501.04248} {arXiv:1501.04248} \BibitemShut
  {NoStop}%
\bibitem [{\citenamefont {Kharel}\ \emph {et~al.}()\citenamefont {Kharel},
  \citenamefont {Behunin}, \citenamefont {Renninger},\ and\ \citenamefont
  {Rakich}}]{Kharel2016}%
  \BibitemOpen
  \bibfield  {author} {\bibinfo {author} {\bibfnamefont {P.}~\bibnamefont
  {Kharel}}, \bibinfo {author} {\bibfnamefont {R.~O.}\ \bibnamefont {Behunin}},
  \bibinfo {author} {\bibfnamefont {W.~H.}\ \bibnamefont {Renninger}}, \ and\
  \bibinfo {author} {\bibfnamefont {P.~T.}\ \bibnamefont {Rakich}},\
  }\href@noop {} {\ }\Eprint {http://arxiv.org/abs/arXiv:1512.07606v1}
  {arXiv:arXiv:1512.07606v1} \BibitemShut {NoStop}%
\bibitem [{\citenamefont {Dragic}(2009)}]{Dragic2009}%
  \BibitemOpen
  \bibfield  {author} {\bibinfo {author} {\bibfnamefont {P.~D.}\ \bibnamefont
  {Dragic}},\ }\href {\doibase 10.1016/j.jnoncrysol.2009.01.005} {\bibfield
  {journal} {\bibinfo  {journal} {Journal of Non-Crystalline Solids}\ }\textbf
  {\bibinfo {volume} {355}},\ \bibinfo {pages} {403} (\bibinfo {year}
  {2009})}\BibitemShut {NoStop}%
\bibitem [{\citenamefont {Park}\ and\ \citenamefont {Wang}(2009)}]{Park2009}%
  \BibitemOpen
  \bibfield  {author} {\bibinfo {author} {\bibfnamefont {Y.-S.}\ \bibnamefont
  {Park}}\ and\ \bibinfo {author} {\bibfnamefont {H.}~\bibnamefont {Wang}},\
  }\href {\doibase 10.1038/nphys1303} {\bibfield  {journal} {\bibinfo
  {journal} {Nature Physics}\ }\textbf {\bibinfo {volume} {5}},\ \bibinfo
  {pages} {489} (\bibinfo {year} {2009})}\BibitemShut {NoStop}%
\bibitem [{\citenamefont {Bahl}\ \emph
  {et~al.}(2011{\natexlab{a}})\citenamefont {Bahl}, \citenamefont
  {Zehnpfennig}, \citenamefont {Tomes},\ and\ \citenamefont
  {Carmon}}]{Bahl2011}%
  \BibitemOpen
  \bibfield  {author} {\bibinfo {author} {\bibfnamefont {G.}~\bibnamefont
  {Bahl}}, \bibinfo {author} {\bibfnamefont {J.}~\bibnamefont {Zehnpfennig}},
  \bibinfo {author} {\bibfnamefont {M.}~\bibnamefont {Tomes}}, \ and\ \bibinfo
  {author} {\bibfnamefont {T.}~\bibnamefont {Carmon}},\ }\href {\doibase
  10.1038/ncomms1412} {\bibfield  {journal} {\bibinfo  {journal} {Nature
  communications}\ }\textbf {\bibinfo {volume} {2}},\ \bibinfo {pages} {403}
  (\bibinfo {year} {2011}{\natexlab{a}})}\BibitemShut {NoStop}%
\bibitem [{\citenamefont {Bahl}\ \emph
  {et~al.}(2011{\natexlab{b}})\citenamefont {Bahl}, \citenamefont {Tomes},
  \citenamefont {Marquardt},\ and\ \citenamefont {Carmon}}]{Bahl2011a}%
  \BibitemOpen
  \bibfield  {author} {\bibinfo {author} {\bibfnamefont {G.}~\bibnamefont
  {Bahl}}, \bibinfo {author} {\bibfnamefont {M.}~\bibnamefont {Tomes}},
  \bibinfo {author} {\bibfnamefont {F.}~\bibnamefont {Marquardt}}, \ and\
  \bibinfo {author} {\bibfnamefont {T.}~\bibnamefont {Carmon}},\ }\href
  {\doibase 10.1038/nphys2206} {\bibfield  {journal} {\bibinfo  {journal}
  {Nature Physics}\ }\textbf {\bibinfo {volume} {8}},\ \bibinfo {pages} {203}
  (\bibinfo {year} {2011}{\natexlab{b}})}\BibitemShut {NoStop}%
\bibitem [{\citenamefont {Anetsberger}\ \emph {et~al.}(2009)\citenamefont
  {Anetsberger}, \citenamefont {Rivi{\`{e}}re}, \citenamefont {Sehliesser},
  \citenamefont {Arcizet},\ and\ \citenamefont {Kippenberg}}]{Anetsberger2009}%
  \BibitemOpen
  \bibfield  {author} {\bibinfo {author} {\bibfnamefont {G.}~\bibnamefont
  {Anetsberger}}, \bibinfo {author} {\bibfnamefont {R.}~\bibnamefont
  {Rivi{\`{e}}re}}, \bibinfo {author} {\bibfnamefont {A.}~\bibnamefont
  {Sehliesser}}, \bibinfo {author} {\bibfnamefont {O.}~\bibnamefont {Arcizet}},
  \ and\ \bibinfo {author} {\bibfnamefont {T.~J.}\ \bibnamefont {Kippenberg}},\
  }\href {\doibase 10.1109/CLEOE-EQEC.2009.5192612} {\bibfield  {journal}
  {\bibinfo  {journal} {Nature Photonics}\ }\textbf {\bibinfo {volume} {2}},\
  \bibinfo {pages} {627} (\bibinfo {year} {2009})}\BibitemShut {NoStop}%
\bibitem [{\citenamefont {Schliesser}\ \emph {et~al.}(2008)\citenamefont
  {Schliesser}, \citenamefont {Rivi{\`{e}}re}, \citenamefont {Anetsberger},
  \citenamefont {Arcizet},\ and\ \citenamefont {Kippenberg}}]{Schliesser2008}%
  \BibitemOpen
  \bibfield  {author} {\bibinfo {author} {\bibfnamefont {A.}~\bibnamefont
  {Schliesser}}, \bibinfo {author} {\bibfnamefont {R.}~\bibnamefont
  {Rivi{\`{e}}re}}, \bibinfo {author} {\bibfnamefont {G.}~\bibnamefont
  {Anetsberger}}, \bibinfo {author} {\bibfnamefont {O.}~\bibnamefont
  {Arcizet}}, \ and\ \bibinfo {author} {\bibfnamefont {T.~J.}\ \bibnamefont
  {Kippenberg}},\ }\href@noop {} {\bibfield  {journal} {\bibinfo  {journal}
  {Nature Physics}\ }\textbf {\bibinfo {volume} {4}},\ \bibinfo {pages} {415}
  (\bibinfo {year} {2008})}\BibitemShut {NoStop}%
\bibitem [{\citenamefont {Rivi{\`{e}}re}\ \emph {et~al.}(2011)\citenamefont
  {Rivi{\`{e}}re}, \citenamefont {Del{\'{e}}glise}, \citenamefont {Weis},
  \citenamefont {Gavartin}, \citenamefont {Arcizet}, \citenamefont
  {Schliesser},\ and\ \citenamefont {Kippenberg}}]{Riviere2011}%
  \BibitemOpen
  \bibfield  {author} {\bibinfo {author} {\bibfnamefont {R.}~\bibnamefont
  {Rivi{\`{e}}re}}, \bibinfo {author} {\bibfnamefont {S.}~\bibnamefont
  {Del{\'{e}}glise}}, \bibinfo {author} {\bibfnamefont {S.}~\bibnamefont
  {Weis}}, \bibinfo {author} {\bibfnamefont {E.}~\bibnamefont {Gavartin}},
  \bibinfo {author} {\bibfnamefont {O.}~\bibnamefont {Arcizet}}, \bibinfo
  {author} {\bibfnamefont {A.}~\bibnamefont {Schliesser}}, \ and\ \bibinfo
  {author} {\bibfnamefont {T.~J.}\ \bibnamefont {Kippenberg}},\ }\href
  {\doibase 10.1103/PhysRevA.83.063835} {\bibfield  {journal} {\bibinfo
  {journal} {Physical Review A}\ }\textbf {\bibinfo {volume} {83}},\ \bibinfo
  {pages} {1} (\bibinfo {year} {2011})}\BibitemShut {NoStop}%
\bibitem [{\citenamefont {Okawachi}\ \emph {et~al.}(2005)\citenamefont
  {Okawachi}, \citenamefont {Bigelow}, \citenamefont {Sharping}, \citenamefont
  {Zhu}, \citenamefont {Schweinsberg}, \citenamefont {Gauthier}, \citenamefont
  {Boyd},\ and\ \citenamefont {Gaeta}}]{Okawachi2005}%
  \BibitemOpen
  \bibfield  {author} {\bibinfo {author} {\bibfnamefont {Y.}~\bibnamefont
  {Okawachi}}, \bibinfo {author} {\bibfnamefont {M.~S.}\ \bibnamefont
  {Bigelow}}, \bibinfo {author} {\bibfnamefont {J.~E.}\ \bibnamefont
  {Sharping}}, \bibinfo {author} {\bibfnamefont {Z.}~\bibnamefont {Zhu}},
  \bibinfo {author} {\bibfnamefont {A.}~\bibnamefont {Schweinsberg}}, \bibinfo
  {author} {\bibfnamefont {D.~J.}\ \bibnamefont {Gauthier}}, \bibinfo {author}
  {\bibfnamefont {R.~W.}\ \bibnamefont {Boyd}}, \ and\ \bibinfo {author}
  {\bibfnamefont {A.~L.}\ \bibnamefont {Gaeta}},\ }\href {\doibase
  10.1103/PhysRevLett.94.153902} {\bibfield  {journal} {\bibinfo  {journal}
  {Physical Review Letters}\ }\textbf {\bibinfo {volume} {94}},\ \bibinfo
  {pages} {1} (\bibinfo {year} {2005})}\BibitemShut {NoStop}%
\bibitem [{\citenamefont {Stokes}\ \emph {et~al.}(1982)\citenamefont {Stokes},
  \citenamefont {Chodorow},\ and\ \citenamefont {Shaw}}]{Stokes1982}%
  \BibitemOpen
  \bibfield  {author} {\bibinfo {author} {\bibfnamefont {L.~F.}\ \bibnamefont
  {Stokes}}, \bibinfo {author} {\bibfnamefont {M.}~\bibnamefont {Chodorow}}, \
  and\ \bibinfo {author} {\bibfnamefont {H.~J.}\ \bibnamefont {Shaw}},\ }\href
  {\doibase 10.1364/OL.7.000509} {\bibfield  {journal} {\bibinfo  {journal}
  {Optics letters}\ }\textbf {\bibinfo {volume} {7}},\ \bibinfo {pages} {509}
  (\bibinfo {year} {1982})}\BibitemShut {NoStop}%
\bibitem [{\citenamefont {Boyd}\ and\ \citenamefont
  {Gauthier}(2009)}]{Boyd2009}%
  \BibitemOpen
  \bibfield  {author} {\bibinfo {author} {\bibfnamefont {R.~W.}\ \bibnamefont
  {Boyd}}\ and\ \bibinfo {author} {\bibfnamefont {D.~J.}\ \bibnamefont
  {Gauthier}},\ }\href {\doibase 10.1126/science.1170885} {\bibfield  {journal}
  {\bibinfo  {journal} {Science}\ }\textbf {\bibinfo {volume} {326}},\ \bibinfo
  {pages} {1074} (\bibinfo {year} {2009})}\BibitemShut {NoStop}%
\bibitem [{\citenamefont {Kim}\ \emph {et~al.}(2015)\citenamefont {Kim},
  \citenamefont {Kuzyk}, \citenamefont {Han}, \citenamefont {Wang},\ and\
  \citenamefont {Bahl}}]{Kim2015}%
  \BibitemOpen
  \bibfield  {author} {\bibinfo {author} {\bibfnamefont {J.}~\bibnamefont
  {Kim}}, \bibinfo {author} {\bibfnamefont {M.~C.}\ \bibnamefont {Kuzyk}},
  \bibinfo {author} {\bibfnamefont {K.}~\bibnamefont {Han}}, \bibinfo {author}
  {\bibfnamefont {H.}~\bibnamefont {Wang}}, \ and\ \bibinfo {author}
  {\bibfnamefont {G.}~\bibnamefont {Bahl}},\ }\href {\doibase
  10.1038/nphys3236} {\bibfield  {journal} {\bibinfo  {journal} {Nature
  Physics}\ }\textbf {\bibinfo {volume} {11}},\ \bibinfo {pages} {275}
  (\bibinfo {year} {2015})}\BibitemShut {NoStop}%
\bibitem [{\citenamefont {Vahala}\ \emph {et~al.}(2009)\citenamefont {Vahala},
  \citenamefont {Herrmann}, \citenamefont {Kn{\"{u}}nz}, \citenamefont
  {Batteiger}, \citenamefont {Saathoff}, \citenamefont {H{\"{a}}nsch},\ and\
  \citenamefont {Udem}}]{Vahala2009}%
  \BibitemOpen
  \bibfield  {author} {\bibinfo {author} {\bibfnamefont {K.}~\bibnamefont
  {Vahala}}, \bibinfo {author} {\bibfnamefont {M.}~\bibnamefont {Herrmann}},
  \bibinfo {author} {\bibfnamefont {S.}~\bibnamefont {Kn{\"{u}}nz}}, \bibinfo
  {author} {\bibfnamefont {V.}~\bibnamefont {Batteiger}}, \bibinfo {author}
  {\bibfnamefont {G.}~\bibnamefont {Saathoff}}, \bibinfo {author}
  {\bibfnamefont {T.~W.}\ \bibnamefont {H{\"{a}}nsch}}, \ and\ \bibinfo
  {author} {\bibfnamefont {T.}~\bibnamefont {Udem}},\ }\href {\doibase
  10.1038/nphys1367} {\bibfield  {journal} {\bibinfo  {journal} {Nature
  Physics}\ }\textbf {\bibinfo {volume} {5}},\ \bibinfo {pages} {682} (\bibinfo
  {year} {2009})}\BibitemShut {NoStop}%
\bibitem [{\citenamefont {Anderson}\ \emph {et~al.}(1972)\citenamefont
  {Anderson}, \citenamefont {Halperin},\ and\ \citenamefont
  {Varma}}]{Anderson1972}%
  \BibitemOpen
  \bibfield  {author} {\bibinfo {author} {\bibfnamefont {P.~W.}\ \bibnamefont
  {Anderson}}, \bibinfo {author} {\bibfnamefont {B.~I.}\ \bibnamefont
  {Halperin}}, \ and\ \bibinfo {author} {\bibfnamefont {C.~M.}\ \bibnamefont
  {Varma}},\ }\href {\doibase 10.1080/14786437208229210} {\bibfield  {journal}
  {\bibinfo  {journal} {Philosophical Magazine}\ }\textbf {\bibinfo {volume}
  {25}},\ \bibinfo {pages} {1} (\bibinfo {year} {1972})}\BibitemShut {NoStop}%
\bibitem [{\citenamefont {Rakich}\ \emph {et~al.}(2012)\citenamefont {Rakich},
  \citenamefont {Reinke}, \citenamefont {Camacho}, \citenamefont {Davids},\
  and\ \citenamefont {Wang}}]{Rakich2012}%
  \BibitemOpen
  \bibfield  {author} {\bibinfo {author} {\bibfnamefont {P.~T.}\ \bibnamefont
  {Rakich}}, \bibinfo {author} {\bibfnamefont {C.}~\bibnamefont {Reinke}},
  \bibinfo {author} {\bibfnamefont {R.}~\bibnamefont {Camacho}}, \bibinfo
  {author} {\bibfnamefont {P.}~\bibnamefont {Davids}}, \ and\ \bibinfo {author}
  {\bibfnamefont {Z.}~\bibnamefont {Wang}},\ }\href
  {http://link.aps.org/doi/10.1103/PhysRevX.2.011008} {\bibfield  {journal}
  {\bibinfo  {journal} {Physical Review X}\ }\textbf {\bibinfo {volume} {2}},\
  \bibinfo {pages} {11008} (\bibinfo {year} {2012})}\BibitemShut {NoStop}%
\bibitem [{\citenamefont {Jen}\ \emph {et~al.}(1986)\citenamefont {Jen},
  \citenamefont {Safaai-Jazi},\ and\ \citenamefont {Farnell}}]{Jen1986}%
  \BibitemOpen
  \bibfield  {author} {\bibinfo {author} {\bibfnamefont {C.~K.}\ \bibnamefont
  {Jen}}, \bibinfo {author} {\bibfnamefont {A.}~\bibnamefont {Safaai-Jazi}}, \
  and\ \bibinfo {author} {\bibfnamefont {G.~W.}\ \bibnamefont {Farnell}},\
  }\href {\doibase 10.1109/T-UFFC.1986.26878} {\bibfield  {journal} {\bibinfo
  {journal} {IEEE Transactions on Ultrasonics, Ferroelectrics, and Frequency
  Control}\ }\textbf {\bibinfo {volume} {33}},\ \bibinfo {pages} {634}
  (\bibinfo {year} {1986})}\BibitemShut {NoStop}%
\bibitem [{\citenamefont {Skacel}\ \emph {et~al.}(2015)\citenamefont {Skacel},
  \citenamefont {Kaiser}, \citenamefont {Wuensch}, \citenamefont {Rotzinger},
  \citenamefont {Lukashenko}, \citenamefont {Jerger}, \citenamefont {Weiss},
  \citenamefont {Siegel},\ and\ \citenamefont {Ustinov}}]{Skacel2015}%
  \BibitemOpen
  \bibfield  {author} {\bibinfo {author} {\bibfnamefont {S.~T.}\ \bibnamefont
  {Skacel}}, \bibinfo {author} {\bibfnamefont {C.}~\bibnamefont {Kaiser}},
  \bibinfo {author} {\bibfnamefont {S.}~\bibnamefont {Wuensch}}, \bibinfo
  {author} {\bibfnamefont {H.}~\bibnamefont {Rotzinger}}, \bibinfo {author}
  {\bibfnamefont {A.}~\bibnamefont {Lukashenko}}, \bibinfo {author}
  {\bibfnamefont {M.}~\bibnamefont {Jerger}}, \bibinfo {author} {\bibfnamefont
  {G.}~\bibnamefont {Weiss}}, \bibinfo {author} {\bibfnamefont
  {M.}~\bibnamefont {Siegel}}, \ and\ \bibinfo {author} {\bibfnamefont {A.~V.}\
  \bibnamefont {Ustinov}},\ }\href@noop {} {\bibfield  {journal} {\bibinfo
  {journal} {Applied Physics Letters}\ }\textbf {\bibinfo {volume} {022603}}
  (\bibinfo {year} {2015})}\BibitemShut {NoStop}%
\end{thebibliography}%

\end{document}